\documentclass[]{interact}

\usepackage[T1]{fontenc}
\usepackage{lmodern}
\usepackage{amsmath,amssymb}
\usepackage{graphicx}
\usepackage{float}
\usepackage{booktabs}
\usepackage{longtable}
\usepackage[numbers,sort&compress,merge]{natbib}
\bibpunct[, ]{[}{]}{,}{n}{,}{,}

\usepackage[colorlinks=true,linkcolor=blue,citecolor=blue,urlcolor=blue]{hyperref}
\hypersetup{
  pdftitle={Liquid-liquid criticality in a fluid whose two structures favor
            mixing},
  pdfauthor={Marcio Pazetti},
  pdfsubject={Liquid-liquid criticality; two-state thermodynamics;
              density-order coupling; excluded volume},
  pdfkeywords={liquid-liquid critical point, two-state thermodynamics,
               non-ideality of mixing, excluded volume, density-order coupling}}

\newcommand{\kb}{k_{\mathrm{B}}}

\newcommand{\TLG}{T^{*}_{c,\mathrm{LG}}}
\newcommand{\ueff}{u_{\rm eff}}

\newcommand{\ORCID}{0000-0002-7567-5470}
\newcommand{\ZenodoDOI}{10.5281/zenodo.23197159}

\begin{document}
\received{Compiled November 10, 2026}

\title{Liquid--liquid criticality in a fluid whose two structures favor mixing}

\author{
\name{Marcio Pazetti\thanks{CONTACT Marcio Pazetti. Email: marcio.pazetti@univasf.edu.br}}
\affil{Colegiado de Engenharia de Produ\c{c}\~ao, Universidade Federal do Vale do S\~ao Francisco,
Av.\ Ant\^onio Carlos Magalh\~aes 510, Country Club, Juazeiro, BA 48902-300, Brazil}
}

\maketitle

\begin{abstract}
The generic two-state description of liquid polyamorphism requires a non-ideality
opposing mixing for liquid--liquid separation. This condition assumes that the free
energy of the underlying fluid is independent of the structural order, an assumption
that fails when the local structures differ in size or attraction. We derive a
closed-form stability condition after density relaxation for a van der Waals closure
based on the second virial coefficient of a pair potential whose internal state
determines both the contact distance and the well depth. The condition recovers the
familiar result for structures identical in size and attraction but can change sign
within the liquid when they differ. In ten of the thirty-four parameter sets
examined, a critical state preferred by the liquid requires a non-ideality favoring
mixing. For one such fluid, we construct the coexistence curve, which terminates at
the predicted critical point at positive pressure. Attraction and excluded volume
provide distinct contributions to the density--order coupling; only the former
survives on a fixed lattice with single occupancy. Removing the excluded-volume
contribution restores the familiar condition in $58$ of the $82$ such critical states
found. Thus, within this closure, liquid--liquid separation can occur with a
non-ideality favoring mixing.
\end{abstract}

\begin{keywords}
liquid--liquid critical point; two-state thermodynamics; non-ideality of mixing;
excluded volume; density--order coupling
\end{keywords}

\section{Introduction}
\label{sec:intro}

A liquid--liquid transition is a transition between two amorphous states of a
one-component fluid. Since the proposal of a second critical point in
supercooled water~\cite{Poole1992} the phenomenon has been reported or inferred
in phosphorus~\cite{Katayama2000}, silicon~\cite{SastryAngell2003},
cerium~\cite{Cadien2013} and water under pressure~\cite{Kim2020}; the
amorphous--amorphous transition of ice~\cite{Mishima1985} is the historical
antecedent and Rapoport's regular-solution treatment of a two-species
liquid~\cite{Rapoport1967} an early thermodynamic account. High-accuracy simulation places the critical point of the DNN@MB-pol water
model at $188\pm5$~K and $1050\pm50$~atm and, correcting for that model's
systematic offset from the experimental phase diagram, infers one for real water
near $198\pm5$~K and $1250\pm50$~atm~\cite{NatPhys2025}, and recent work reports molecular
evidence for two interconverting local structures in liquid water~\cite{Li2026}.

Isotropic
potentials with a softened core or two characteristic length scales have been
known for two decades to produce liquid--liquid transitions and second critical
points~\cite{Franzese2001}. Two-state thermodynamics~\cite{Anisimov2018}
organizes the phenomenology at a different level, writing
\begin{equation}
G=G_A+xG_{BA}+\kb T\big[x\ln x+(1-x)\ln(1-x)\big]+\omega\,x(1-x),
\label{eq:aniG}
\end{equation}
with $x$ the fraction in the second structure, $G_{BA}=\Delta H-T\Delta S+p\,\Delta V$
the field, $\Delta H$ being the enthalpy change at zero pressure, and $\omega$ the
non-ideality, two roles that formulation already
distinguishes. Closer still to the present work,
Ref.~\cite{Anisimov2025} shows that \emph{symmetric} interconversion, in which
the two structures are degenerate and no field distinguishes them, can produce
tricriticality once the interconversion is coupled to density. Our symmetric
sector reproduces that mechanism inside a second-virial closure, and
Sec.~\ref{sec:sym2} states what is specific to it.

This work connects the two levels. Starting from a pair potential, the one of
Eq.~\eqref{eq:pot} below, it determines how much of Eq.~\eqref{eq:aniG} a
second-virial closure fixes, and where in the closure's parameter space a critical
point appears once density and internal order are treated as coupled fluctuations. We use the algebraic structure of the closure of Ref.~\cite{SPD2001}, a van der Waals
equation for a Lennard-Jones attraction outside a hard core, that is, a Sutherland
potential~\cite{GilVillegas1997}, whose contact distance and well depth are modulated by the relative
orientation of the molecules. That closure belongs to the generalized van der
Waals treatment of anisotropic fluids, in which a hard core and a mean-field attractive
tail are both modulated by a particle coordinate~\cite{GelbartBaron1977,Wensink2009}; the line that
carries a Sutherland tail on a modulated core runs from Ref.~\cite{TeloDaGama1984}, with a
spherical core, through Ref.~\cite{Teixeira1999}, which lets the contact distance depend on
the coordinate, to Ref.~\cite{SPD2001}, and is reviewed in Ref.~\cite{FrancoMelgar2009}.
What changes here is the coordinate itself: it is an internal two-state variable rather than
an orientation, so the ordering it produces is liquid--liquid rather than nematic.
Earlier work on the van der Waals equation
obtained liquid--liquid criticality by perturbing it beyond the second virial
coefficient~\cite{SPYP2020}; here we ask what the second-virial closure achieves
on its own. The second virial coefficient enters only as interaction input, fixing
the excluded volume and the attraction; the free energy built from them has the van der
Waals form, whose excluded-volume term contains all powers of the density, so the closure
describes liquid densities, and the critical points found in Sec.~\ref{sec:region} lie at reduced volumes
between $1.00$ and $2.02$ times the excluded volume.

\section{The closure}
\label{sec:closure}

\subsection{Interaction, moments, units}

The interaction is
\begin{equation}
u(r,H)=
\begin{cases}
+\infty, & r<r_0\,[1-\gamma\alpha H],\\[4pt]
-\mu_0\,[1+\alpha H]
\left\{\dfrac{r_0[1-\gamma\alpha H]}{r}\right\}^{6}, & r\ge r_0\,[1-\gamma\alpha H],
\end{cases}
\label{eq:pot}
\end{equation}
with $\alpha$ the first-order response of the well depth to the internal state and
$\gamma\alpha$ that of the contact distance, so that $\gamma$ is the ratio of the two
and the core amplitude is the product $s\equiv\gamma\alpha$. The excluded volume goes
as the cube of the contact distance, so its own first-order response is $3\gamma\alpha$
--- which is why the moments of Eq.~\eqref{eq:moments} carry $(1-sH)^{3}$ and not
$(1-sH)$. In Ref.~\cite{SPD2001} these arise as
derivatives at zero of a continuous orientational coordinate; a derivative at
zero is not a property of a binary variable, so here they are to be read as
effective coefficients inherited from that parent model and used as the defining
parameters of the present one.
Writing $\langle\cdot\rangle$ for the average over the distribution of $H$,
\begin{equation}
C_0=\big\langle(1-sH)^3\big\rangle,\quad
C_1=\big\langle(1-sH)^3(1+\alpha H)\big\rangle,\quad
C_2=\big\langle(1-sH)^3(1+\alpha H)^2\big\rangle ,
\label{eq:moments}
\end{equation}
the equation of state is of van der Waals form with $b=V_mC_0/8$, where $V_m=4\pi r_0^3/3$, and
$a(T)=(V_m\mu_0/8)[C_1+(\beta\mu_0/6)C_2]$. The factor $8$ belongs to the
unnormalized orientational measure implied by the polynomials of
Ref.~\cite{SPD2001}, which give $C_0=4$ at $\alpha=0$, that is, $\langle1\rangle=4$; with the normalized two-state measure used from
Sec.~\ref{sec:sym2} onwards the same excluded volume reads
$b=V_m\langle A^3\rangle/2$, and both give $b=V_m/2$ in the isotropic limit. We
verified Eq.~\eqref{eq:moments} term by term against the polynomials of
Ref.~\cite{SPD2001} (Appendix~\ref{app:moments}); the full normalization chain,
with the check $B_2^{\rm core}\to2\pi r_0^3/3$, is in Appendix~\ref{app:norm}.
From here $\kb=1$, energy is in units of $\mu_0$ and volume in units of $V_m$.

Throughout, $\TLG$ denotes the liquid--gas critical temperature of the
\emph{underlying closure}, computed at $m=0$ with neither field nor
non-ideality. It is a reference scale, not the liquid--gas critical temperature
of the extended model that any particular critical point below would require.

\subsection{Two-state coordinate and the symmetric sector}
\label{sec:sym2}

Let $\sigma=\pm1$ label the two local structures and $m=\langle\sigma\rangle$.
This is not the closure of Ref.~\cite{SPD2001} evaluated with a different
measure, and it should not be described as one. Three operations separate them.
First, the coordinate modulating the contact distance and the well depth is
orientational there and continuous, with an angular measure; here it is a
discrete two-state variable, with the state measure and mixing entropy of a
discrete degree of freedom. Second, the one-particle term
$\kappa(\sigma_1+\sigma_2)/2$ in the contact factor has no counterpart in the
parent model; it is introduced here, and it is what makes the excluded volume
depend on a particle's own state. Third, Ref.~\cite{SPD2001} truncates the
expansion of the second virial coefficient in $\beta$ at the three moments
$C_0$, $C_1$ and $C_2$, whereas Eq.~\eqref{eq:allorders} resums the Mayer series
of the attractive tail to all orders in $\beta$, with the interaction parameters
still obtained from the two-body cluster integral alone. The resulting free energy
is not a literal truncation at $O(\rho^{2})$: the term $-T\ln[v-b(m)]$ of
Eq.~\eqref{eq:free} is a van der Waals resummation of the excluded volume and
contains all powers of density when expanded. What is of second virial order is
the interaction input, not the free energy. The second-virial treatment of an anisotropic
pair interaction goes back to Ref.~\cite{Onsager1949}. Its known weakness is at liquid
densities, where a free energy truncated at $O(\rho^{2})$ omits the higher virial
terms and is customarily repaired by Parsons--Lee scaling~\cite{Parsons1979,Lee1987},
which is also how algebraic equations of state for nematics are built on Onsager's
theory~\cite{FrancoMelgar2008}; that scaling has no target here, since the excluded volume enters through the resummed
logarithmic term rather than through a truncated series.

What is inherited is the algebraic structure: the interaction is resolved
through three moments of the form of Eq.~\eqref{eq:moments}, and the equation of
state retains the van der Waals form with state-dependent $b$ and $a$. What is native to the
present construction is the thermodynamic mapping onto a discrete two-state
coordinate, the density--order coupling that follows from it, and the extraction
of the critical non-ideality. The model is therefore a two-state closure built
on that structure, and the consequences that occupy the rest of this paper ---
the coupling of density to internal order, and the sign of the non-ideality it
requires --- belong to it rather than to Ref.~\cite{SPD2001}. Because $\sigma=\pm1$,
the average over two independent particles is a sum over four configurations with
weights $[1+(\sigma_1+\sigma_2)m+\sigma_1\sigma_2m^2]/4$, so every average is
exactly quadratic in $m$; hence $b(m)$ and $a(T,m)$ are quadratic and the Mayer
series can be summed without truncation,
\begin{equation}
a(T,m)=\frac{T}{2}\Big\langle A^{3}\,\hat S(B/T)\Big\rangle,\qquad
\hat S(z)=\sum_{n\ge1}\frac{z^{n}}{n!\,(2n-1)},
\label{eq:allorders}
\end{equation}
with $A=1-s\,\sigma_1\sigma_2+\kappa(\sigma_1+\sigma_2)/2$ and
$B=1+\alpha\,\sigma_1\sigma_2$. Written out, the pair potential actually solved
here is the two-state Sutherland form
\begin{equation}
u_{\sigma_1\sigma_2}(r)=
\begin{cases}
\infty, & r<r_0A(\sigma_1,\sigma_2),\\[2pt]
-\mu_0\,B(\sigma_1,\sigma_2)\big[r_0A(\sigma_1,\sigma_2)/r\big]^{6}, & r\ge r_0A(\sigma_1,\sigma_2),
\end{cases}
\label{eq:pot2state}
\end{equation}
so that the internal states set both the contact distance, through $A$, and the
well depth, through $B$. An analogous construction with the coordinate taken to be an
orientation is the generalized van der Waals nematic of
Ref.~\cite{Teixeira1999}, which took as its starting point the spherical-core model of Ref.~\cite{TeloDaGama1984}, where the
orientation sets only the attraction, and made the contact distance depend on the
orientation as well. Two-particle averages are taken with
$P(\sigma_1,\sigma_2)=P(\sigma_1)P(\sigma_2)$ and $P(\sigma)=(1+\sigma m)/2$,
which is the mean-field assumption of the treatment and the only one made about
correlations between internal states.  The one-particle term in $\kappa$ lets the core of a particle depend
on its own state, breaking the parity in $m$. 

Two formal distinctions must be established here. Equation~\eqref{eq:allorders} is exact in two senses
that should not be confused. The average over internal states is exact because
$\sigma=\pm1$ makes it a sum over four configurations, not an expansion; and
$\hat S$ is the exact sum of the Mayer series of the attractive tail at fixed
contact distance, $\int_{r_0}^{\infty}\!\big[e^{-\beta u}-1\big]r^2dr$ reduced by
term-by-term integration of $e^{-\beta u}-1$ for $u\propto r^{-6}$, each term
giving the factor $1/(2n-1)$. What is not exact is the second-virial input
itself, which is a property of the closure and not of Eq.~\eqref{eq:allorders};
and $\hat S$ is an infinite series, evaluated numerically here.

The free energy per particle that all later derivatives are taken of is
\begin{equation}
f(T,v,m)=-T\ln\!\big[v-b(m)\big]-\frac{a(T,m)}{v}
+T\Big[x\ln x+(1-x)\ln(1-x)\Big],\qquad x=\frac{1+m}{2},
\label{eq:free}
\end{equation}
up to terms depending on $T$ alone, which do not enter any derivative used
below. The field and the non-ideality enter through the Hamiltonian of the
two-state fluid of $N$ particles,
\[
\mathcal{H}=\sum_i\frac{\Delta H}{2}\,\sigma_i
+\sum_{i<j}\Big[u_{\sigma_i\sigma_j}(r_{ij})-\frac{\omega'}{2}\,\sigma_i\sigma_j\Big].
\]
Here $\sigma_i=\pm1$ is the structure of particle $i$, with $\sigma=+1$ the second
structure of Eq.~\eqref{eq:aniG}; $\Delta H$ is the energy of the second structure
relative to the first; $u_{\sigma\sigma'}$ is the pair potential of
Eq.~\eqref{eq:pot2state} and $r_{ij}$ the distance between particles $i$ and $j$;
and $\omega'=\omega/N$ is the strength with which the structures of any two
particles are coupled. That coupling has the Curie--Weiss form: it is the same
for every pair, whatever their separation, so it carries no length scale and
does not depend on $v$, and the factor $1/N$ is what keeps its total energy
proportional to $N$. The same infinite-range form appears in the double-occupancy
cell model of Ref.~\cite{Romanik2026}, where it acts between cell occupancies and,
together with an in-cell repulsion, produces a liquid--liquid transition classified
by density alone; here it acts between internal structures. That infinite range is the physical content of the
assumption. The entropy difference does not appear in $\mathcal{H}$, which does
not depend on $T$; it is carried by the degeneracies $g_\pm$ of the two
structures, $\ln(g_+/g_-)=\Delta S$. In the thermodynamic limit the mean-field
factorization stated above is exact for the coupling term, which contributes
$-\omega m^{2}/4=\omega x(1-x)-\omega/4$ per particle, while the degeneracies
contribute $-(m/2)T\Delta S$. Dropping the constant,
$f\to f+\tfrac{m}{2}(\Delta H-T\Delta S)+\omega\,x(1-x)$. The constant $\omega$
is a parameter of $\mathcal{H}$ that the pair potential does not fix, and it may
take either sign.

Positivity of the contact distance over all four configurations requires
$A>0$. Over the domain of parameter signs actually sampled here --- $\alpha>0$,
$\gamma>0$, $\kappa>0$ --- the binding constraint is
\begin{equation}
A_{\min}=1-s-\kappa>0 ,
\label{eq:Amin}
\end{equation}
attained at $\sigma_1=\sigma_2=-1$; for other sign choices the minimum may fall
on a different configuration. Since a parameter set with $A_{\min}$ near zero has
one internal configuration whose contact distance nearly collapses, we require
$A_{\min}>0.3$ throughout and report $A_{\min}$ with every parameter set of Tables~\ref{tab:interval} and~\ref{tab:states} and of the Supplemental Material. This excludes
eleven of the $66$ combinations of the grids of Table~\ref{tab:grids}, leaving
$55$. The value $0.3$ is a regularity choice, not a physical constant. Repeating the
mapping under six other cuts, from $0.15$ to $0.45$, the fraction of admissible sets
that admit solutions stays between $0.649$ and $0.833$, the largest value at the tightest cut
(Appendix~\ref{app:grids}), so the region is not manufactured by the cut,
though the cut does bound the domain searched; the grids are identified in
Table~\ref{tab:grids}.

\paragraph*{The symmetric sector.}
At $\kappa=0$ the free energy is even in $m$, so at $m=0$ one has $f_{vm}=0$: the
Hessian is diagonal, the soft mode is the internal coordinate $\mathbf{e}=(0,1)$,
the stable mode is the density $\mathbf{n}=(1,0)$ with $\lambda_n=f_{vv}$, and
$D^3f[\mathbf{e},\mathbf{e},\mathbf{e}]=f_{mmm}=0$ identically. Criticality is
then reached by choosing $\omega=2f_{mm}$, and what decides the order of the
transition is the \emph{relaxed} quartic of Eq.~\eqref{eq:ueff} below, which in
this sector reads
\begin{equation}
\ueff^{(\kappa=0)}=f_{mmmm}-\frac{3f_{mmv}^{2}}{f_{vv}} .
\label{eq:ueffsym}
\end{equation}
This is Eq.~\eqref{eq:ueff} at $\kappa=0$: there $f_{vm}=0$ at $m=0$, so
$\mathbf{e}=(0,1)$ and $\mathbf{n}=(1,0)$, whence
$H_\omega[\mathbf{n},\mathbf{n}]=f_{vv}$ and the $-\tfrac{\omega}{2}n_m^{2}$ term is
absent.
The cross term does not vanish, because $b$ and $a$ depend on $m^2$, so
$\ueff$ is not the quartic at fixed volume.

The symmetric sector therefore carries no Schur correction to the quadratic
coefficient, and what density fluctuations renormalize there is the quartic.
That is the mechanism of Ref.~\cite{Anisimov2025}, where tricriticality is obtained
for symmetric interconversion coupled to density in mean field and Monte Carlo
and is rationalized by a Landau expansion in which the coupling renormalizes the
quartic coefficient [their Appendix~A, Eq.~(14)]; the symmetric sector of this
closure reproduces it, and in it the coupling is not introduced as a separate phenomenological term but
follows from $b=V_m\langle A^3\rangle/2$ with $A$ depending on the internal
state. The sector serves here as the control against which the rest of the
paper is read: it is where the bilinear coupling vanishes by symmetry. Its
consequences for the order of the symmetric transition are a separate question
and are not pursued here.

\section{Criticality in a compressible fluid}
\label{sec:crit}

Since $\kappa\neq0$ makes $b'(0)\neq0$, we have $f_{vm}\neq0$: internal
order and density are coupled and the mode that softens is a mixture of the two.
At fixed $(T,p)$ the relevant potential is
$\Phi=f+\tfrac{m}{2}(\Delta H-T\Delta S)+\omega\tfrac{1-m^{2}}{4}+pv$, whose Hessian in
$(v,m)$ we write $H_\omega\equiv D^2f-\tfrac{\omega}{2}\,\hat e_m\hat e_m^{\mathsf T}$,
and criticality requires
\begin{equation}
\boxed{\;
f_{vv}\Big(f_{mm}-\tfrac{\omega}{2}\Big)-f_{vm}^{2}=0,\qquad
D^3f[\mathbf{e},\mathbf{e},\mathbf{e}]=0,\qquad
\ueff>0,\qquad f_{vv}>0 , \;}
\label{eq:crit}
\end{equation}
with $\mathbf{e}$ spanning the null space of $H_\omega$ and $\mathbf{n}$ any direction
independent of it.  Imposing
$f_m=f_{mm}=f_{mmm}=0$ at
fixed $v$ would be the criticality of a system of fixed density, which is not the
problem a liquid poses. The quartic condition is taken after relaxing the stable mode,
\begin{equation}
\ueff=D^4f[\mathbf{e},\mathbf{e},\mathbf{e},\mathbf{e}]
-\frac{3\,\big(D^3f[\mathbf{e},\mathbf{e},\mathbf{n}]\big)^{2}}{H_\omega[\mathbf{n},\mathbf{n}]},
\label{eq:ueff}
\end{equation}
since $D^4f[\mathbf{e}]>0$ alone can classify as stable a point whose effective
quartic is negative --- and Sec.~\ref{sec:sym2} shows that the subtracted term does not vanish in the symmetric sector. Written this way
$\ueff$ does not depend on the choice of $\mathbf{n}$: at a point satisfying the first two
conditions above, $H_\omega\mathbf{e}=0$ gives $H_\omega[\mathbf{e},\mathbf{n}]=0$ for
every $\mathbf{n}$, and $D^3f[\mathbf{e},\mathbf{e},\mathbf{e}]=0$ removes the component
of $\mathbf{n}$ along $\mathbf{e}$, so replacing $\mathbf{n}$ by $a\mathbf{e}+b\mathbf{n}$
multiplies numerator and denominator alike by $b^{2}$. In the implementation
$\mathbf{e}\propto(-f_{vm},f_{vv})$ normalized and $\mathbf{n}=(-e_2,e_1)$, which is a
convenience and not a requirement. What the choice of $\mathbf{e}$ does fix is scale:
$\ueff$ is homogeneous of degree four in $\mathbf{e}$, so its sign, which is what
classifies, is independent of the normalization, while the magnitudes quoted in
Table~\ref{tab:states} are those of $\|\mathbf{e}\|=1$ in the
Euclidean metric of $(v/V_m,m)$. Mechanical stability is not optional: a stationary point at which $f_{vv}\le0$
cannot represent a locally mechanically stable liquid critical state.

Neither the field nor $\omega\,x(1-x)$ enters $f_{vv}$, $f_{vm}$ or any third or
fourth derivative, the first being linear in $m$ and the second quadratic and
$v$-independent. It is therefore useful to isolate the internal curvature that
remains once density has been relaxed and before either is included,
\begin{equation}
R_{\rm cl}\equiv f^{(0)}_{mm}-\frac{\big(f^{(0)}_{vm}\big)^{2}}{f^{(0)}_{vv}},
\qquad\text{so that}\qquad \omega_c=2R_{\rm cl} ,
\label{eq:omegac}
\end{equation}
the subtracted term being the softening of the internal mode by its coupling to
density, and $R_{\rm cl}$ being the Schur complement of the Hessian with respect
to the density mode. We call it the \emph{relaxed baseline curvature} rather than
the contribution of the closure alone, because it is not only that: it contains
the curvature of the ideal mixing entropy of the two-state degree of freedom.

The scope of Eq.~\eqref{eq:omegac} should be stated plainly, because it is easy
to overclaim. It is a \emph{structural identity}: for any free energy of a
two-state coordinate carrying a non-ideality $\omega\,x(1-x)$, eliminating a
second mode by mechanical equilibrium produces the Schur complement, and
marginality of the surviving mode reads $\omega_c=2R_{\rm cl}$. Nothing in the
derivation uses the Sutherland tail, the second-virial truncation or the specific
form of $b(m)$. The identity is therefore not a result of this closure and we do
not present it as one. What the closure supplies, and what the rest of this paper
is about, is the explicit function $R_{\rm cl}(T,v,m;\alpha,\gamma,\kappa)$
below: where in the liquid region it is positive, where it vanishes, where it
becomes negative, and which of those states survive the remaining criticality
conditions and the test of global stability.
Writing $\xi=v-b(m)$ and primes for $\partial_m$, the derivatives are
\begin{gather}
f_{mm}=\frac{T}{1-m^{2}}+T\!\left[\frac{b''}{\xi}+\frac{(b')^{2}}{\xi^{2}}\right]-\frac{a''}{v},
\qquad
f_{vm}=-\frac{Tb'}{\xi^{2}}+\frac{a'}{v^{2}},
\qquad
f_{vv}=\frac{T}{\xi^{2}}-\frac{2a}{v^{3}},
\end{gather}
so that
\begin{equation}
R_{\rm cl}=\underbrace{\frac{T}{1-m^{2}}}_{\text{ideal}}
+\underbrace{T\!\left[\frac{b''}{\xi}+\frac{(b')^{2}}{\xi^{2}}\right]-\frac{a''}{v}
-\frac{f_{vm}^{2}}{f_{vv}}}_{\text{induced by the closure}} .
\label{eq:rclsplit}
\end{equation}
The first term is the ideal reference and survives at infinite dilution; the
second is what the closure adds. Two things follow that are easy to get wrong.
The ideal part is $T/(1-m^{2})$, not $T$: it reduces to $T$ only at $m=0$.
And $b'$ and $a'$ vanish at $m=0$ only in the symmetric sector: with
$\kappa\neq0$ the linear coefficient $b_1$ is nonzero, so $b'(0)\neq0$, while at
$\kappa=0$ one has $b'=2b_2m$ and $a'=2a_2m$, so $f_{vm}$, proportional to $m$, is already
$4.9\times10^{-3}$ at $m=0.01$ (at $\alpha=0.05$, $\gamma=0.50$, $T=0.75\,\TLG$ and
$v=1.505\,b_0$, where $\ueff$ vanishes on the liquid branch at $m=0$). At $m=0$ the Hessian is diagonal, $f_{vm}=0$
exactly, and the closure-induced coupling drops out, so the compact form that one might
write from the quadratic coefficients alone,
\begin{equation}
R_{\rm cl}\big|_{m=0,\,\kappa=0}=T+\frac{2Tb_2}{\xi}-\frac{2a_2}{v},
\end{equation}
carries no $-f_{vm}^2/f_{vv}$ term at all, and must not be used elsewhere. All numerical results below use
Eq.~\eqref{eq:rclsplit} in full. What Eq.~\eqref{eq:omegac} fixes is the non-ideality
\emph{required at a prescribed thermodynamic state}, not $\omega$ itself, and its
sign carries physical content. Since $\omega x(1-x)=\omega(1-m^2)/4$ contributes $-\omega/2$
to the curvature in $m$, positive $\omega$ penalizes mixing and promotes
separation while negative $\omega$ promotes mixing; neither is excluded. Three
regimes follow:
\begin{itemize}
\item $R_{\rm cl}>0$: the closure is internally subcritical and a
demixing-promoting non-ideality, $\omega_c>0$, is required. In the dilute limit
the closure contributes nothing and $R_{\rm cl}\to T/(1-m^{2})$, so that
$Q\equiv(1-m^{2})\omega_c/2T\to1$; only at $m=0$ does this reduce to the
regular-solution condition $\omega_c/2T\to1$, and quoting the latter away from
$m=0$ is the error corrected in Sec.~\ref{sec:compare}. At liquid densities the
closure renormalizes that threshold, upwards
or downwards depending on the state, and over part of the state space carries it
through zero into the third regime below.
\item $R_{\rm cl}=0$: the baseline is exactly at internal marginality and
$\omega_c=0$.
\item $R_{\rm cl}<0$: the closure has already overshot the internal instability,
and a mixing-promoting non-ideality, $\omega_c<0$, is required to bring the state
back to marginality.
\end{itemize}
 Retaining only roots with $\omega_c>0$ would be a modeling
convention, not a thermodynamic requirement, and Sec.~\ref{sec:region} shows how
much of the solution set it would discard.

Stationarity is part of the definition of a critical point and fixes the
remaining fields. A root of Eq.~\eqref{eq:crit} at $(T,v,m)$ determines
$p_c=-f_v$ from $\Phi_v=0$, and the field required for that state to be
stationary in the internal coordinate follows from $\Phi_m=0$, that is
$h_c\equiv\Delta H-T\Delta S=\omega_c m_c-2f_m$. The root fixes $p_c$, $\omega_c$
and the field together; it does not fix $\Delta H$ and $\Delta S$ separately,
which are one-particle inputs.

Equation~\eqref{eq:omegac} admits a second derivation, shorter than the Hessian
route and closer to the two-state description itself, which we give because it
identifies what $\omega_c$ is. Eliminate the volume along the mechanically
equilibrated liquid branch $v_{\rm liq}(m)$ defined by $f_v+p_c=0$ and form the
constrained Gibbs energy of the extended model, Eq.~\eqref{eq:gibbs} below.
Because $f_v+p_c$ vanishes identically along that branch, the term in
$dv_{\rm liq}/dm$ cancels and
\begin{equation}
g'(m)=f_m+\frac{h_c}{2}-\frac{\omega_c m}{2},
\qquad
g''(m)=f_{mm}-\frac{f_{vm}^{2}}{f_{vv}}-\frac{\omega_c}{2}
=R_{\rm cl}-\frac{\omega_c}{2},
\label{eq:gderivs}
\end{equation}
where $g''$ picks up $-f_{vm}^2/f_{vv}$ through
$dv_{\rm liq}/dm=-f_{vm}/f_{vv}$, the same quotient that defines the soft
eigenvector. Substituting $h_c=\omega_c m_c-2f_m$ and $\omega_c=2R_{\rm cl}$ makes
$g'(m_c)$ and $g''(m_c)$ vanish \emph{identically}. Equation~\eqref{eq:omegac} is
therefore not merely the value of $\omega$ that closes the Hessian determinant:
it is exactly the non-ideality for which $m_c$ is a degenerate stationary point of
the constrained Gibbs energy of the two-state model, which is what a
liquid--liquid critical point in the internal coordinate means. Both residuals
were also evaluated numerically over the roots found and are at round-off,
$|g'(m_c)|<7.2\times10^{-15}$ and $g''(m_c)=0$ exactly
(Appendix~\ref{app:num}).

\section{Where solutions exist}
\label{sec:region}

\subsection{Search and observable}

The search is restricted to the mechanically stable liquid branch and
parametrized by $\xi=v-b(m)$, logarithmically from $10^{-4}$, so the
neighborhood of $v\to b$ is resolved for every $m$; we write $\xi$ rather than
$x$ to avoid collision with the structural fraction $x=(1+m)/2$ of
Eq.~\eqref{eq:aniG}. For each $m$ the branch ends at the smallest $\xi$ with
$f_{vv}<0$. Solutions are retained if $A>0$, $B>0$, $|m|<1$, $f_{vv}>0$ and
$\ueff>0$, and Eq.~\eqref{eq:allorders} is used throughout. No condition is
imposed on the sign of $\omega_c$.
The presence counts below are taken over the $34$ triples $(\alpha,\gamma,\kappa)$
of the main grid with $A_{\min}>0.3$, each searched at $0.80$ and $0.92\,\TLG$; the
grids are identified in Table~\ref{tab:grids} of Appendix~\ref{app:grids}.

The solutions of Eq.~\eqref{eq:crit} form continuous branches in $m$. The
search sweeps a mesh in $m$ and, along each line of constant $m$, locates the roots
in $\xi$, so each mesh line samples those branches and the number of roots found
grows with the number of lines; neither that number nor any fraction built from it
is a property of the closure, and we report neither as a result (Appendix~\ref{app:conv}).
What is invariant is the presence or absence of a given kind of solution in a
given parameter set. Those counts are reported throughout, and
Table~\ref{tab:audit} gives them; they are unchanged under refinement of the mesh
in $m$, except that the two global counts settle only at $136$ values
(Appendix~\ref{app:conv}). The symmetric sector of Sec.~\ref{sec:sym2} is a
different object: there $m=0$ is fixed by symmetry and there is no mesh in $m$ at
all.

For temperatures we report the interval over which at least one solution
exists, with both edges located by bisection between a sampled temperature with a
solution and the adjacent one without. The bisection bounds only that outer loop.
The inner test --- does a solution exist at this temperature --- rests on finite
meshes in $m$ and $\xi$, and the two directions do not behave alike: a coarse mesh
in $\xi$, the direction along which roots are bracketed, can fail to bracket a sign
change at all and drop a root from the count. All results below whose output is a
count or the position of a boundary therefore use $400$ geometric values of $\xi$, densified as in Appendix~\ref{app:mesh}, at which
the presence counts are stationary, and convergence is assessed in each mesh
direction separately (Appendix~\ref{app:conv}). We treat differences between
parameter sets smaller than about $0.03$ in $T/\TLG$ as not quantitatively
interpretable, a conservative threshold given the convergence of $T_+$ reported in
Sec.~\ref{sec:locus}.

\subsection{The boundary $T_+$ of the states requiring positive non-ideality}
\label{sec:locus}

The observable we report is a temperature, and it is important to state exactly
which one, because two descriptions of it that might be taken as independent are in fact
the same. We take as the primary quantity $T_+$, the
lower boundary in temperature of the states that require $\omega_c>0$, that is,
the lowest temperature at which a root with $\omega_c>0$ is found at all. A
second description is often used: the temperature at which the largest
$\omega_c$ over the roots found changes sign; Fig.~\ref{fig:rcl}(a) displays the
equivalent envelope in $Q$, for the reason given below. Since
$\max_{\rm roots}\omega_c>0$ holds if and only if some root has $\omega_c>0$,
the two coincide wherever the second exists. The two descriptions are not interchangeable
in general, and Table~\ref{tab:interval} shows why: in two of the nine sets the
critical solutions cease to be found while $\omega_c$ is still positive
throughout, so $T_+$ exists as the end of the region carrying positive roots
while no sign change of the envelope is observed. We therefore report $T_+$, and
say for each set whether a sign change accompanies it. It is not the locus $R_{\rm cl}=0$, which would require solving
$R_{\rm cl}=0$ simultaneously with the remaining criticality conditions over the
whole branch; we have not done that, and the two should not be conflated.

What follows from $T_+$ is asymmetric, and Fig.~\ref{fig:rcl}(c) shows why:
at a single temperature the closure can carry roots with $\omega_c>0$ and roots
with $\omega_c<0$ at once. Above $T_+$ all that follows is that \emph{at
least one} root requires $\omega_c>0$; others at the same temperature may require
the opposite sign. Below it, every root detected requires $\omega_c<0$, and that
statement is complete over the sampled population. The substantive content is
therefore not that the boundary exists but what lies below it. In seven of the
nine sets examined in detail, removing the sign restriction reveals locally
critical states below $T_+$, and every state detected there requires
$\omega_c<0$; in the remaining two the solutions cease at $T_+$ while $\omega_c$
is still positive.

A second distinction matters as much. The roots found at different temperatures
are not, in general, critical points of one fluid. Each carries its own
$\omega_c$ and its own one-particle field, so a set of roots is a family of
extended models. The correct statement is that the closure continues to admit
locally critical states below $T_+$, with those states requiring
mixing-promoting non-ideality; whether any particular fluid remains critical
there is a separate question, answered only by fixing
$(\omega,\Delta H,\Delta S)$ as in Sec.~\ref{sec:binodal}.

A third concerns how the envelope changes sign. It can do so because the root
carrying the maximum passes continuously through $\omega_c=0$, or because the
branch of positive roots is lost from the sampled set, leaving an unrelated root
to carry the maximum below. Tracking the maximizing root across $T_+$ on the
converged mesh, the first happens in six of the seven sets in which a sign
change occurs: the same root carries the maximum on both sides, with $m$ and $v/b(m_c)$
fixed to the mesh resolution and $\omega_c$ small and of opposite signs. In
$(\alpha,\gamma,\kappa)=(0.10,0.50,0.20)$ the positive root instead holds
$m\simeq+0.058$ with $\omega_c$ falling only from $+0.468$ at $0.750\,\TLG$ to
$+0.459$ at $T_+$ and is then no longer found, the maximum below being carried by a
root at $m=-0.902$ with $\omega_c=-5.15$. We call this a \emph{detected} termination of the positive
branch rather than a termination, because the search locates roots by sign
changes of $D^3f$ on a finite mesh and a tangential root of even multiplicity
would be missed; establishing termination as such would require continuation of
the branch, which we have not carried out. Table~\ref{tab:interval} records which
of the two each set is. Figure~\ref{fig:rcl} plots the envelope of
$Q=(1-m^{2})\omega_c/2T$ rather than of $\omega_c$, for the reason given in
Sec.~\ref{sec:compare}; since $1-m^{2}>0$ the two envelopes change sign at the
same temperature, so $T_+$ is unaffected, and we checked that the maximizing root
is also the same under either measure at the two sides of $T_+$ in six of the seven
sets and, in the seventh, $(0.10,0.50,0.20)$ above $T_+$, a neighboring root of the
same window, $0.002$ away in $m$, so the classification above is unchanged.

The first three presence counts of Table~\ref{tab:audit} are invariant under
refinement of the mesh in $m$, and the two global counts settle at $136$ values;
$T_+$, however, is the position of a boundary and needs a test of its own, which we ran at fixed $\xi=400$ for all nine sets. For seven of the nine, refining the
mesh in $m$ from $40$ to $68$ to $136$ values moves $T_+$ by at most $0.0021$,
against a bisection tolerance of $10^{-4}$, and we report it to three decimals.
The remaining two, $(0.10,0.50,0.20)$ and $(0.20,0.50,0.40)$, are not resolved
by a uniform mesh at all: near $T_+$ the states requiring $\omega_c>0$ survive
only in a window in $m$ narrower than any uniform step, so the value returned
drifts downward as the mesh is refined instead of settling. For those two the
boundary is located by refining the mesh in $m$ locally, which gives $0.745$
and $0.583$.

The counts we report are counts of parameter sets, and they are
these. Of the $34$ parameter sets of Table~\ref{tab:audit}, $29$ admit at least one locally
critical root; $21$ admit at least one requiring $\omega_c>0$ and $20$ at least
one requiring $\omega_c<0$; and, under the global-stability filter of
Sec.~\ref{sec:binodal}, $12$ admit at least one globally preferred state
requiring $\omega_c>0$ and $10$ at least one requiring $\omega_c<0$. These
categories are not mutually exclusive: twelve sets carry roots of both signs,
which is why $21+20$ exceeds $29$.
The states requiring mixing-promoting non-ideality are therefore not a marginal
population: they occur in nearly as many parameter sets as the states requiring
the opposite sign, and are globally preferred in almost as many.

\begin{table}[htbp]
\centering
\caption{%
Presence of each kind of critical state across the $34$ parameter sets
with $A_{\min}>0.3$, at two temperatures each, $0.80$ and $0.92\,\TLG$, with
order-parameter meshes of $136$ and $271$ values, which agree, and $400$ values
of $\xi$. A set is counted once if at least one root of the stated kind was
found in it. The columns do not partition the sets: twelve carry roots of both
signs.  Convergence under refinement of both meshes is reported in
Appendix~\ref{app:conv}. The set-by-set listing from which these totals are taken is given in
the Supplemental Material; absolute numbers of roots are not reported as results, for the
reason given in Sec.~\ref{sec:region}.
}
\label{tab:audit}
{\small\setlength{\tabcolsep}{5pt}
\begin{tabular*}{\columnwidth}{@{\extracolsep{\fill}}ccccc}
\toprule
any root & $\omega_c>0$ & $\omega_c<0$ &
global, $\omega_c>0$ & global, $\omega_c<0$\\
\midrule
29 & 21 & 20 & 12 & 10\\
\bottomrule
\end{tabular*}}
\end{table}

\begin{table}[htbp]
\centering
\caption{%
The lower boundary $T_+$ of the states requiring $\omega_c>0$, in units
of $\TLG$, on the converged mesh and with no condition on the sign of $\omega_c$
anywhere in the search. Values are quoted to three decimals. For seven of the nine the spread between
$40$, $68$ and $136$ values of $m$ is at most $0.0021$. The printed values, like
that sequence of meshes, are located by bisection to a bracket no wider than
$10^{-4}$ whose two ends round to the same third decimal; a plain bisection to
$5\times10^{-4}$ gives the same third decimal for all nine.
For $(0.10,0.50,0.20)$ and $(0.20,0.50,0.40)$ the roots
that survive nearest the boundary lie within $0.02\%$ of the liquid spinodal and
occupy a window in $m$ narrower than any uniform step, so uniform meshes return
values that drift downward with refinement; $0.745$ and $0.583$ are obtained by
refining the mesh in $m$ locally. The last column says whether a sign change of the upper envelope
accompanies the boundary and, if so, by which mechanism, obtained by tracking the
maximizing root across it: \emph{continuous} means the same root carries the
maximum on both sides and passes through zero; \emph{detected termination} means
the positive branch is no longer found below, the maximum being carried by an
unrelated root. In two sets no sign change is seen: there the critical
solutions cease to be found while $\omega_c$ is still positive, so $T_+$ is the
temperature at which they cease and not a sign change.
}
\label{tab:interval}
{\small\setlength{\tabcolsep}{5pt}
\begin{tabular*}{\textwidth}{@{\extracolsep{\fill}}cccccll}
\toprule
$\alpha$ & $\gamma$ & $\kappa$ & $s$ & $A_{\min}$ & $T_+/\TLG$ & sign change of the envelope\\
\midrule
0.10 & 0.50 & 0.20 & 0.05 & 0.750 & 0.745 & yes, detected termination\\
0.10 & 0.50 & 0.40 & 0.05 & 0.550 & 0.695 & none; solutions cease with $\omega_c>0$\\
0.20 & 0.50 & 0.20 & 0.10 & 0.700 & 0.887 & yes, continuous\\
0.20 & 0.50 & 0.30 & 0.10 & 0.600 & 0.694 & yes, continuous\\
0.20 & 0.50 & 0.40 & 0.10 & 0.500 & 0.583 & none; solutions cease with $\omega_c>0$\\
0.20 & 0.70 & 0.30 & 0.14 & 0.560 & 0.794 & yes, continuous\\
0.20 & 0.70 & 0.40 & 0.14 & 0.460 & 0.586 & yes, continuous\\
0.30 & 0.50 & 0.40 & 0.15 & 0.450 & 0.832 & yes, continuous\\
0.15 & 1.40 & 0.30 & 0.21 & 0.490 & 0.843 & yes, continuous\\
\bottomrule
\end{tabular*}}
\end{table}

\begin{figure}[htbp]
\centering
\includegraphics[width=\linewidth]{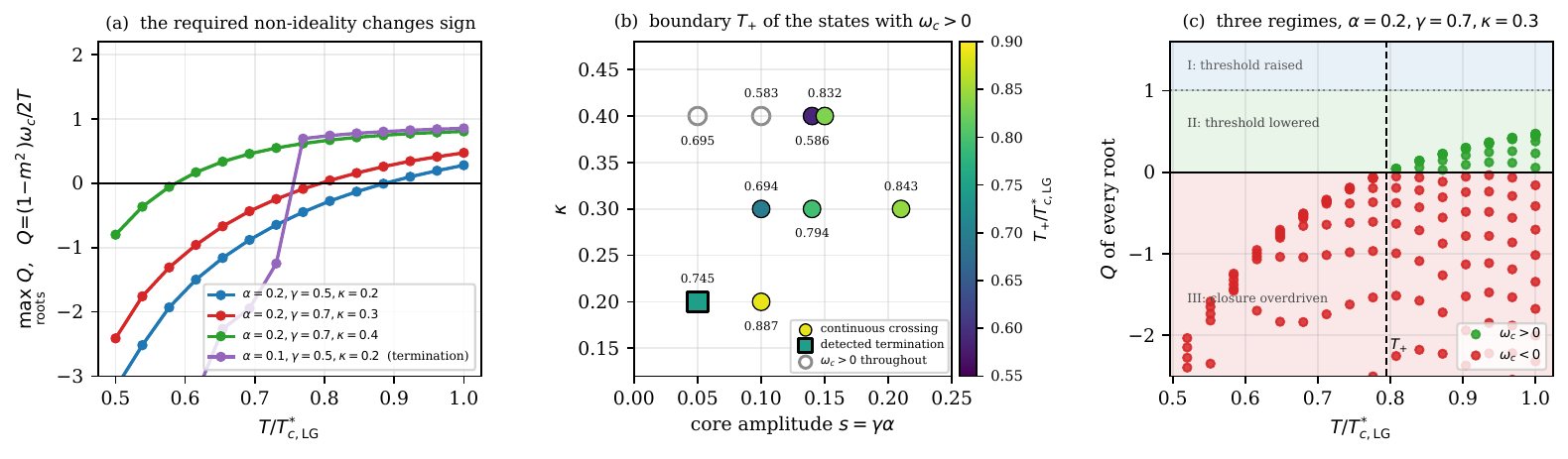}
\caption{%
(a) Upper envelope over the critical roots of
$Q=(1-m^{2})\omega_c/2T$, the required non-ideality referred to the local ideal
value of Sec.~\ref{sec:compare}, against temperature and with no condition on the
sign of $\omega_c$, on the converged mesh of $136$ values of $m$ and $400$ of $\xi$,
refined locally in $m$ for $(0.10,0.50,0.20)$ as in Table~\ref{tab:interval}; its
zero crossing, where it occurs, marks $T_+$. We plot $Q$
rather than $\omega_c/2T$ because for $m\neq0$ the ideal reference is
$1/(1-m^{2})$ and not unity, so a comparison with the regular-solution value would
otherwise be made against the wrong number. (b) The boundary $T_+$ of the states requiring
$\omega_c>0$ in the plane of core amplitude and one-particle asymmetry; open symbols are the two sets
in which the solutions cease while $\omega_c$ is still positive
(Table~\ref{tab:interval}), and the set marked
by a square is the one in which the envelope changes sign by detected termination
of the positive branch rather than by a continuous crossing (Table~\ref{tab:interval}).
This is not the locus $R_{\rm cl}=0$, which would require solving that condition
simultaneously with the criticality conditions of Eq.~\eqref{eq:crit}. (c) Every root of one parameter set on the same mesh, again in $Q$:
roots of both signs occur at the same temperature, which is why the statement
above $T_+$ is existential and the one below it universal. The shaded bands are
$Q>1$, where the closure raises the threshold, $0<Q<1$, where it lowers it, and
$Q<0$, where it has overshot.
}
\label{fig:rcl}
\end{figure}

\subsection{The upper edge}
\label{sec:upper}

The upper edge of the temperature range carrying critical roots sits just above
the reference scale in the nine sets of Table~\ref{tab:interval}, between $1.025$
and $1.092\,\TLG$ on the converged mesh, and is unaffected by the condition on
$\omega_c$ in eight of them; in two sets an isolated root with $\omega_c<0$
reappears above that edge, up to $1.15\,\TLG$.
It is close to $\TLG$ but not equal to it, and the two are different objects:
$\TLG$ is defined at $m=0$ without field or non-ideality, whereas the upper edge
is where the states with $m\neq0$ are losing their mechanically stable liquid branch.
Over the same range, for $(\alpha,\gamma,\kappa)=(0.20,0.50,0.30)$, the number of $m$ values possessing a liquid branch falls
from all $68$ sampled values at $0.95\,\TLG$ to $40$ at $\TLG$ and none at
$1.20\,\TLG$, so
the range terminates where the liquid branch is disappearing rather than at an
independent limit of liquid--liquid criticality.

\subsection{Representative critical states}

Table~\ref{tab:states} gives, for one root of each of the nine parameter sets of
Table~\ref{tab:interval}, the quantities needed to judge whether the solutions sit in the interior
of the liquid region; where a set has more than one root at the temperature
quoted we list, among the roots requiring $\omega_c>0$, the one of smallest
$|m|$. Ties in $|m|$ occur, and the tie-break is given in
Appendix~\ref{app:mesh}. The reduced volumes run from $1.12$ to $1.65$ in $v/b(m_c)$, so every state satisfies $v>b(m_c)$ and lies in
the dense liquid, the closest being $12.5\%$ above the excluded-volume bound. Four of the nine have $p_c<0$: those critical points
lie in the stretched, mechanically metastable liquid, which is legitimate but
must be stated. Pressures are in units of $\mu_0/V_m$ and
$\omega_c$ in units of $\mu_0$.

\begin{table}[htbp]
\centering
\caption{%
One representative critical state for each of the nine parameter sets
used in the detailed analysis of $T_+$ in Table~\ref{tab:interval}, taken at the
midpoint between $T_+$ as reported there and $\TLG$. Eight belong to the main grid and one,
$(\alpha,\gamma,\kappa)=(0.15,1.40,0.30)$, to the gap-filling grid. The reduced volume is
$v/b(m_c)$, the excluded volume evaluated at the critical order parameter, which
is the quantity that the domain of Eq.~\eqref{eq:free} requires to exceed unity;
$v/b_0$, with $b_0=b(0)$, is a different quantity and can fall below unity without any state being
inadmissible. In the symmetric
sector of Sec.~\ref{sec:sym2} the two coincide, since $m_c=0$ gives $b(m_c)=b_0$.
Pressures are in $\mu_0/V_m$ and $\omega_c$ in $\mu_0$. Negative $p_c$ marks a
critical point in the stretched liquid, mechanically metastable against
cavitation. $Q=(1-m_c^{2})\,\omega_c/2T$ is the ratio referred to the local ideal
reference of Sec.~\ref{sec:compare} and equals unity when the closure contributes
nothing; $\omega_c/2T$ is shown beside it so that the size of the
$(1-m_c^{2})$ factor can be read off directly.
}
\label{tab:states}
{\small\setlength{\tabcolsep}{5pt}
\begin{tabular*}{\textwidth}{@{\extracolsep{\fill}}ccccccccccc}
\toprule
$\alpha$ & $\gamma$ & $\kappa$ & $A_{\min}$ & $T/\TLG$ & $v/b(m_c)$ & $m_c$ & $p_c$ & $\omega_c/2T$ & $Q$ & $\ueff$\\
\midrule
0.10 & 0.50 & 0.20 & 0.750 & 0.8725 & 1.648 & -0.016 & -0.013 & 0.770 & 0.770 & 0.310\\
0.10 & 0.50 & 0.40 & 0.550 & 0.8475 & 1.125 & -0.016 & $\phantom{-}$2.988 & 1.743 & 1.742 & 0.324\\
0.20 & 0.50 & 0.20 & 0.700 & 0.9435 & 1.402 & $\phantom{-}$0.331 & $\phantom{-}$0.301 & 0.010 & 0.009 & 0.841\\
0.20 & 0.50 & 0.30 & 0.600 & 0.8470 & 1.346 & $\phantom{-}$0.016 & $\phantom{-}$0.267 & 0.227 & 0.227 & 0.073\\
0.20 & 0.50 & 0.40 & 0.500 & 0.7915 & 1.230 & -0.016 & $\phantom{-}$0.601 & 0.610 & 0.610 & 0.237\\
0.20 & 0.70 & 0.30 & 0.560 & 0.8970 & 1.533 & $\phantom{-}$0.363 & -0.038 & 0.331 & 0.287 & 0.051\\
0.20 & 0.70 & 0.40 & 0.460 & 0.7930 & 1.444 & $\phantom{-}$0.268 & -0.215 & 0.707 & 0.656 & 0.014\\
0.30 & 0.50 & 0.40 & 0.450 & 0.9160 & 1.473 & $\phantom{-}$0.236 & $\phantom{-}$0.050 & 0.205 & 0.194 & 0.313\\
0.15 & 1.40 & 0.30 & 0.490 & 0.9215 & 1.530 & $\phantom{-}$0.615 & -0.040 & 0.428 & 0.267 & 1.331\\
\bottomrule
\end{tabular*}}
\end{table}

\subsection{$T_+$ and the parameters of the closure}
\label{sec:lower}

$T_+$ falls as $\kappa$ grows along lines of constant core amplitude, but
that correlation does not identify $\kappa$ as the quantity responsible, and
Eq.~\eqref{eq:Amin} shows why: at fixed $s$, raising $\kappa$ lowers $A_{\min}$
by the same amount, so the two are collinear by construction. Holding $A_{\min}$
fixed instead and moving along $s=1-A_{\min}-\kappa$, with $\alpha=0.20$ and
$\gamma=s/\alpha$, the values at $A_{\min}=0.55$ are $0.826$, $0.604$ and
$0.779$ for $\kappa=0.30$, $0.35$ and $0.40$. What disappears along that
trajectory is the \emph{monotone} trend; a variation of $0.779-0.604=0.175$
remains, which is comparable to the $0.21$ spanned at constant $s=0.14$ over the same range of $\kappa$. The
locus therefore depends on where in the $(s,\kappa)$ plane one moves, but does
not decrease monotonically with $\kappa$ once proximity to core collapse is held
fixed.

Roots occur over $s$ from $0.035$ to $0.600$ on the main grid at the two temperatures of
the grid. The search locates roots, not the connectivity of the region they occupy;
tangential roots of even multiplicity, which sign changes cannot detect, were
searched for directly (Appendix~\ref{app:conv}). We also tested, and
reject, the hypothesis that its boundary is a line of constant $s+\kappa$: the values
without roots, $0.470$--$0.660$, form a subinterval of those with roots,
$0.235$--$0.660$, so $s+\kappa$ does not separate the two classes at all. At those two
temperatures the five sets without roots are exactly those with $\gamma=1.50$; this is a
property of the temperatures and not of the sets, since, on the mesh of $136$ values of $m$, three of the five carry roots at
other temperatures: all three between $0.95$ and $1.00\,\TLG$, where the liquid branch
ceases to be distinct, and two of them also at isolated temperatures from $0.89$ to
$0.94\,\TLG$. The pairs $(0.30,2.00,\kappa)$ and
$(0.40,1.50,\kappa)$, for $\kappa=0.02$ and $0.06$, share $s$, $\kappa$ and $A_{\min}$
and still differ in outcome, so none of those three fixes it; and because $\gamma=1.50$
is sampled only at $\alpha\ge0.30$ and $\gamma=2.00$ only at $\alpha=0.30$, the grid
attributes the outcome to the pair $(\alpha,\gamma)$, not to either separately.

\subsection{Truncation}

Repeating the search with the $O(\beta^{2})$ truncation of the second virial
coefficient, that is, with the three moments $C_0$, $C_1$ and $C_2$ in place of
Eq.~\eqref{eq:allorders}, yields solutions over the same absolute temperature grid, but at
substantially different reduced temperatures, and the truncation misplaces $\TLG$ itself by $12$ to $15\%$ for these parameters. We report only untruncated
results. Establishing a one-to-one correspondence of critical branches between
the two treatments would require a homotopy with the roots tracked, which we have
not carried out.

\section{Coexistence lines with fixed parameters}
\label{sec:binodal}

A root of Eq.~\eqref{eq:crit} is a critical point of a particular extended model.
To exhibit one such model completely we take the root at $\alpha=0.10$,
$\gamma=0.50$, $\kappa=0.20$, read off $\omega=\omega_c=0.67329$ and, choosing
$\Delta S=0$ so that the field is temperature independent,
$\Delta H=0.51533$. These are then held fixed: with $\omega$ at $0.67329$ the
critical point is solved again, which gives $m_c=-0.015764$ and fixes
$\Delta H=0.5153262$, the value used in Table~\ref{tab:coex}. Eliminating the volume by solving
$p(v,m)=p$ on the liquid branch gives the constrained Gibbs energy
\begin{equation}
g(m)=f\big(T,v_{\rm liq}(T,p,m),m\big)+\tfrac{m}{2}\Delta H
+\omega\tfrac{1-m^2}{4}+p\,v_{\rm liq} .
\label{eq:gibbs}
\end{equation}
Because $m$ is not conserved, the equilibrium state is the global minimum of
$g$, and coexistence means two degenerate global minima; we bracket it by the
discontinuity of $\arg\min_m g$ as $p$ is varied and locate it from $\Delta g=0$ between the two liquid minima (Appendix~\ref{app:coex}). At the critical point of that model every residual of the criticality
conditions sits at machine precision, and $f_{vv}=1.4019$ and $\ueff=0.4835$ are
both positive (Appendix~\ref{app:coex}).

The resulting coexistence line terminates at the predicted critical point,
$T_c=0.930\,\TLG$, $m_c=-0.0158$, $v_c=0.847$, $p_c=0.148$, with residual
$|\Delta g|$ below $3\times10^{-14}$ at every point of the line
[Fig.~\ref{fig:binodal}]. The order-parameter gap closes with the mean-field
exponent; since a mean-field functional must give $\beta=1/2$ at an ordinary
critical point, that fit tests the construction rather than the physics and we
report it in Appendix~\ref{app:num} rather than here.

The same construction in the other sign sector is the one that matters, because
the sign of $\omega$ is where this closure departs from the usual two-state
picture~\cite{Anisimov2018}, and because a non-ideality that favors mixing is the last thing one
would expect to support demixing. Taking a state that is a global minimum and
requires mixing-promoting non-ideality --- $\alpha=0.10$, $\gamma=0.50$,
$\kappa=0.30$, at $T_c=0.920\,\TLG$, the one of largest $\omega_c$ among the
critical roots of that set with $\omega_c<0$ (Appendix~\ref{app:num}) --- and fixing $\omega=-1.05296$,
$\Delta H=2.60964$ and $\Delta S=0$, the coexistence line again terminates at the
predicted critical point, with the same mean-field closing of the gap. Its
critical pressure is positive, $p_c=+0.15483$, so unlike four of the
representative states of Table~\ref{tab:states} it does not sit in the stretched
liquid. Table~\ref{tab:coex} lists the coexisting states of both models.

\begin{figure}[htbp]
\centering
\includegraphics[width=\linewidth]{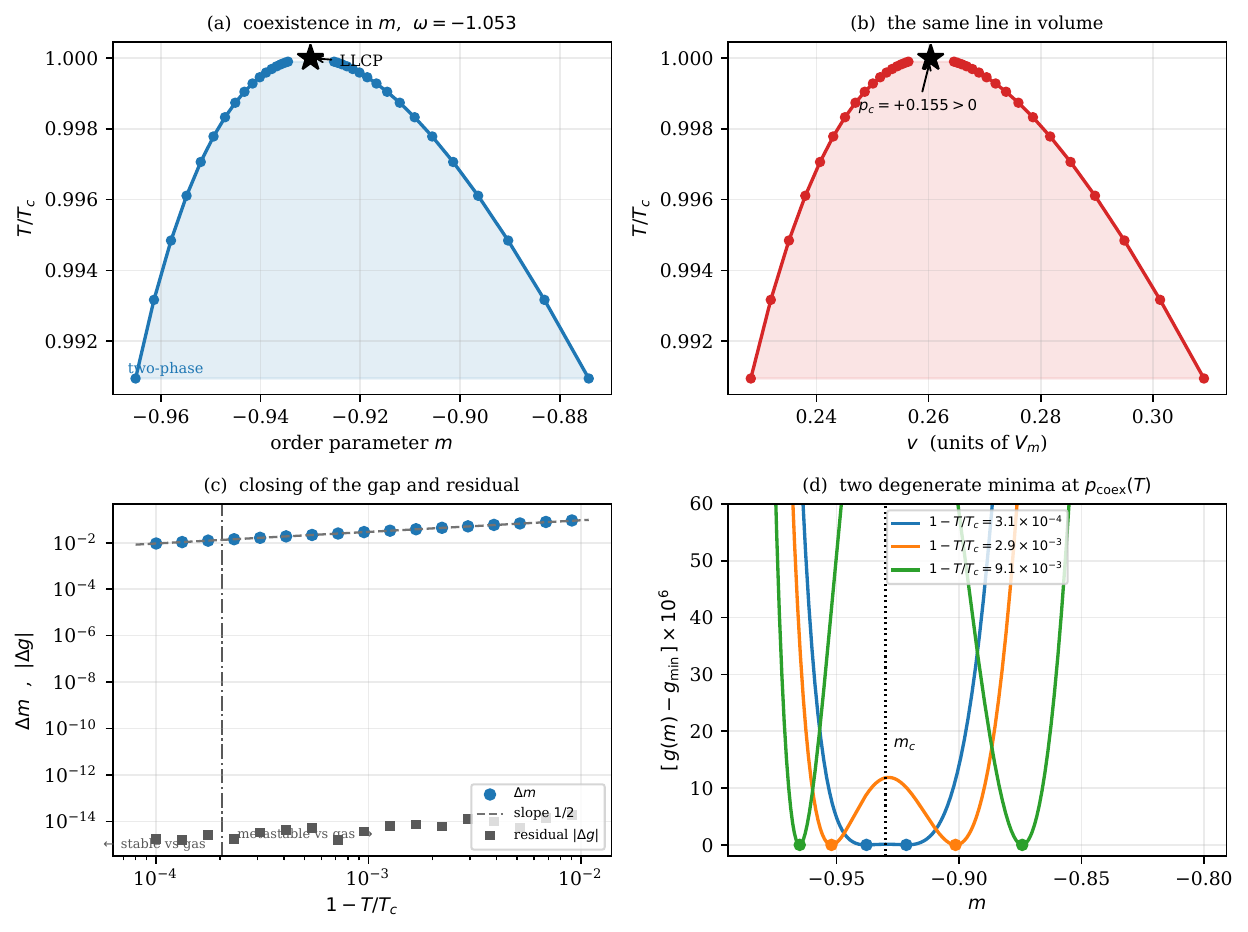}
\caption{%
The fixed-parameter model whose non-ideality favors mixing:
$\alpha=0.10$, $\gamma=0.50$, $\kappa=0.30$, $\omega=-1.05296$,
$\Delta H=2.60964$, $\Delta S=0$, with $T_c=0.920\,\TLG$ and $p_c=+0.15483$.
(a) The coexistence line in the internal order parameter, terminating at the
liquid--liquid critical point. (b) The same line in volume; the critical pressure
is positive, so this transition does not lie in the stretched liquid. (c) The
order-parameter gap against reduced temperature, with a line of slope $1/2$ drawn
for comparison and the dash-dotted line marking $1-T/T_c=2.04\times10^{-4}$, to
the left of which the coexistence is lower in free energy than the vapor and to
the right of which it is metastable against vaporization, and the residual $|\Delta g|$ between the two minima at the
located pressure, which bounds how well the degeneracy is resolved; the fitted
exponent is reported in Appendix~\ref{app:num} rather than here, since a
mean-field functional must give $1/2$. (d) The constrained Gibbs energy along the
liquid branch at the coexistence pressure of three temperatures, showing the two
degenerate minima opening on opposite sides of $m_c$ as the temperature falls.
The line carries seventeen points, spanning $1-T/T_c$ from $10^{-4}$ to
$9.05\times10^{-3}$; each is a coexistence located from $\Delta g=0$ between two
minima that are verified to be interior to their brackets and to have positive
curvature, with $|\Delta g|$ below $2\times10^{-14}$ throughout. Every point is
generated by the scripts in the archived package.
}
\label{fig:neg}
\end{figure}

\begin{figure}[htbp]
\centering
\includegraphics[width=\linewidth]{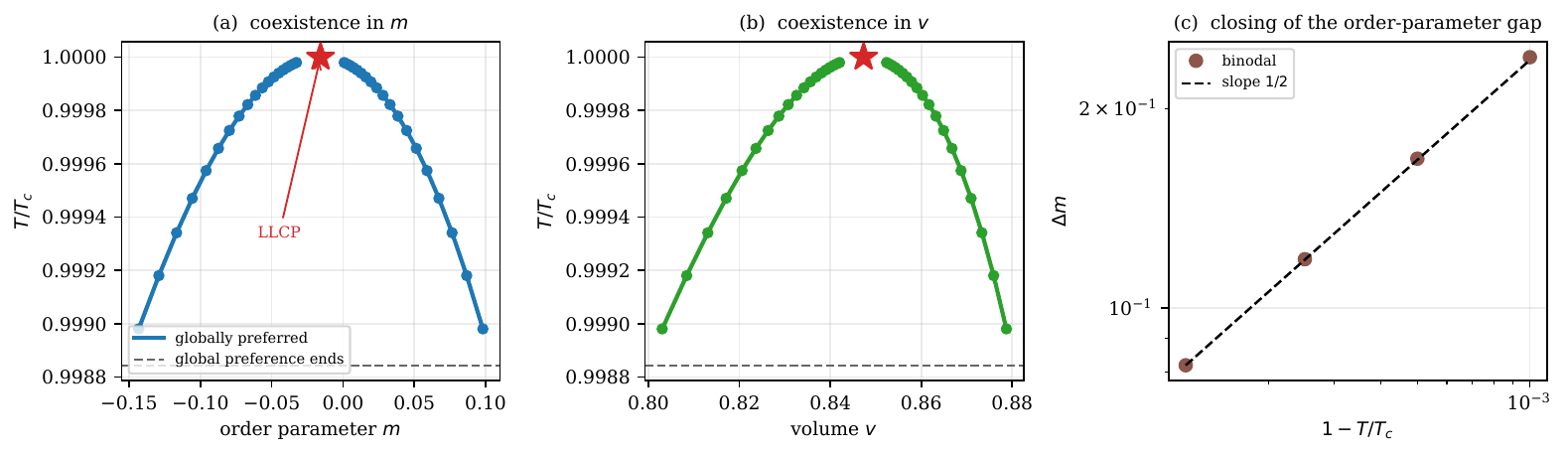}
\caption{%
Liquid--liquid coexistence for the model with $\omega=0.67329$,
$\Delta H=0.51533$, $\Delta S=0$ held fixed, at $\alpha=0.10$, $\gamma=0.50$,
$\kappa=0.20$. \textbf{All three panels show only the liquid--liquid branch},
which the tracking of Appendix~\ref{app:num} follows from the critical point; the
nineteen points plotted in panels (a) and (b), which reach $1-T/T_c=1.02\times10^{-3}$, are exactly those
classified there as globally preferred liquid--liquid coexistence, which ends at
$1.16\times10^{-3}$ (the pair survives as a metastable coexistence up to
$\simeq7\times10^{-3}$, Sec.~\ref{sec:binodal}). Beyond that range the pair of
minima returned by a coexistence search is no longer the liquid--liquid pair ---
its lower branch, at $m$ from $-0.944$ where it takes over to $-0.955$ at
$1-T/T_c=7\times10^{-3}$, is the dense phase --- and the
gap it measures is a liquid--dense one. We keep the two apart rather than
continuing a single curve through them. (a) Coexisting values of the order
parameter and (b) of the volume, with the critical point starred; the dashed line
marks where the liquid--liquid coexistence ceases to be globally preferred, which
we do not claim as a demonstrated triple point. (c) The order-parameter gap
against reduced distance from the critical point, with a line of slope $1/2$ for
comparison; the fitted exponent is discussed in Appendix~\ref{app:num}.
}
\label{fig:binodal}
\end{figure}

This is the central construction of the paper, and Fig.~\ref{fig:neg} shows it.
It exhibits a fully specified fixed-parameter model whose non-ideality favors
mixing and which nevertheless undergoes a liquid--liquid transition, because the closure supplies
the demixing on its own through $R_{\rm cl}$: the coupling of interconversion to
density is strong enough that the mixing term must oppose it for the state to be
merely marginal. The coexistence line is traced over
$1-T/T_c$ from $10^{-4}$ to $9\times10^{-3}$ and closes with the mean-field
exponent. Within the liquid manifold it is the preferred state along its whole
length, the denser competitor that preempts the model at $\omega>0$ after
$0.1\%$ of $T_c$ never taking over here.

Against the gas the statement is different, and we state it because ``globally
preferred'' has meant, throughout this paper, preferred within the liquid
manifold. Enumerating the mechanical roots gives the gas branch as well as the
liquid one, and comparing the constrained Gibbs energy of the two at the
coexistence pressure shows that the liquid is the lower only for
$1-T/T_c<2.04\times10^{-4}$, that is within $0.02\%$ of $T_c$; below that
temperature the vapor has the lower Gibbs energy and the liquid--liquid
transition is metastable against vaporization. This does not remove the
transition, and metastable liquid--liquid transitions are familiar in this
literature, the liquid--liquid critical point of supercooled water lying in a
region metastable against crystallization~\cite{Science2026}; but it must not be read as a
transition between globally stable phases. For the model at $\omega>0$ the
question does not arise in the same form: at its coexistence pressures a mechanically
stable gas root exists only for $m\lesssim-0.92$, where its Gibbs energy lies at
least $0.085$ above the liquid minimum, so no vapor state competes. Panel (d) shows the two minima of $g$ opening on opposite sides of
$m_c$ as the temperature falls, which is the landscape behind the line. The
globally preferred critical points of the sector $\omega_c<0$ on the mesh of $271$ values of $m$ are all strongly ordered, with $m_c$ between
$-0.93$ and $-0.79$ or between $0.64$ and $0.92$, as one expects of an overdriven baseline.

\begin{table}[htbp]
\centering
\caption{%
Coexisting states of the two fixed-parameter models, at
$1-T/T_c=10^{-3}$. Note that this temperature lies below the window
$1-T/T_c<2.04\times10^{-4}$ within which the model at $\omega<0$ is stable against
the vapor, so the second row describes the liquid--liquid coexistence in its
metastable extension; the first row is not affected, since at its coexistence pressures a distinct gas root exists only for
$m\lesssim-0.92$ and lies at least $0.085$ above the liquid minimum of $g$. Pressures are equal at coexistence by construction,
$P_1=P_2=p_{\rm coex}$, and are quoted in $\mu_0/V_m$; densities are
$\rho_i=1/v_i$ in units of $V_m^{-1}$. The pressure is located as the root of
$\Delta g(p)=0$ between the two tracked liquid branches, so the residual
$|\Delta g|$ quoted in the last column is the value of the function being solved
at the returned pressure, and measures how well the degeneracy is resolved. The
two models are specified by $\omega=+0.67329$ with $\Delta H=0.5153262$ and
$\omega=-1.052956$ with $\Delta H=2.609640$, held fixed; the states are
evaluated at $T=T_c(1-\varepsilon)$ with $\varepsilon=10^{-3}$ exactly, which is
not a node of the mesh used for the coexistence lines of
Figs.~\ref{fig:binodal} and~\ref{fig:neg}. 
}
\label{tab:coex}
{\small\setlength{\tabcolsep}{5pt}
\begin{tabular*}{\textwidth}{@{\extracolsep{\fill}}lcccccccc}
\toprule
model & $P_1=P_2$ & $m_1$ & $m_2$ & $v_1$ & $v_2$ & $\rho_1$ & $\rho_2$ & $|\Delta g|$\\
\midrule
$\omega=+0.67329$ & $0.14863$ & $+0.097$ & $-0.142$ & $0.8785$ & $0.8035$ & $1.1383$ & $1.2445$ & $3\times10^{-17}$\\
$\omega=-1.05296$ & $0.14951$ & $-0.914$ & $-0.944$ & $0.2742$ & $0.2483$ & $3.6472$ & $4.0269$ & $2\times10^{-16}$\\
\bottomrule
\end{tabular*}}
\end{table}

Two features of the construction at $\omega>0$ bound the claim made for
\emph{that} model, and we state them here to avoid their being read as properties
of the negative-$\omega$ model of Fig.~\ref{fig:neg}.

First, in the model at $\omega=+0.67329$ the coexistence ceases to be the
preferred one within the liquid manifold at $1-T/T_c\simeq1.16\times10^{-3}$,
where a third, much denser liquid with $m\simeq-0.944$ takes over, and the
liquid--liquid pair itself survives, as a coexistence that is metastable with
respect to that dense phase, to $1-T/T_c\simeq7\times10^{-3}$. We locate the first of these as the sign change of
$g_{\rm LL}-g_{\rm dense}$ at the pressure of liquid--liquid coexistence, which
brackets it at $1.1556\times10^{-3}$, and the second as the last temperature at
which the two liquid minima still become degenerate, bracketed between
$7.0\times10^{-3}$ and $7.2\times10^{-3}$. We do
\emph{not} call the first a triple point: we were unable to exhibit a single
$(T,p)$ at which three minima of Eq.~\eqref{eq:gibbs} are simultaneously
degenerate, so the three-phase character is inferred from a crossing in Gibbs
energy and not demonstrated.

Second, and more consequential, requiring the critical point to be the
\emph{global} minimum of Eq.~\eqref{eq:gibbs} rather than a local one is a strong
filter, and it is what separates a map of local criticality from a map of stable
criticality. Applying it uniformly across the parameter grid of Table~\ref{tab:grids} ---
$34$ parameter sets with $A_{\min}>0.3$, two temperatures each at $0.80$ and
$0.92\,\TLG$ --- gives the presence counts of Table~\ref{tab:audit}. Those come
from a single unrestricted search, with the sign of $\omega_c$ evaluated at each
root and applied afterwards as a filter, so there is one population and not two;
maintaining a separate search with the sign condition imposed from the start
would, we expect, produce apparent inconsistencies of accounting that are artifacts of the
comparison. Because it uses two
temperatures rather than the whole range, this is a sampling of parameter space and not a
complete map of stable criticality. Under the sign condition the sets carrying a
globally preferred state have $\alpha$ between $0.10$ and $0.20$, $s$ between
$0.035$ and $0.140$ and $\kappa$ between $0.20$ and $0.40$, and no set with
$\alpha\ge0.30$ contributes one. That boundary is partly a consequence of the
sign condition: without it the sets at $\alpha=0.30$ acquire states they did not
have, and four of the ten sets carrying a globally preferred state of $\omega_c<0$ lie
outside the region just described --- all four at $\alpha=0.30$. The
quantity that separates the sampled classes most clearly is $\alpha$ rather
than $s$. The competing state is the nearly fully
ordered one, near $\sigma_1=\sigma_2=-1$, where $A=A_{\min}$ and the excluded volume
is smaller by $(A_{\min}/\langle A^3\rangle^{1/3})^3$; the closer $A_{\min}$ is
to zero, the more strongly we expect it to compete, which ties the stability filter to the
same quantity as the positivity cut.

The word ``global'' must be read with care. Equation~\eqref{eq:gibbs} is built
after solving $p(v,m)=p$ on the liquid branch, so the minimum is global in $m$
within the liquid manifold, not over all thermodynamic branches; the gas branch
does not enter this test. For the two fixed-parameter models of
Sec.~\ref{sec:binodal} the gas branch is enumerated as well, and the outcome is
reported there. Four of the states in Table~\ref{tab:states} have
$p_c<0$ and are therefore in the stretched liquid to begin with. Throughout we
write ``globally preferred'' in this sense and mean nothing stronger.

\section{Comparison with two-state thermodynamics}
\label{sec:compare}

The generic formulation of fluid polyamorphism writes the Gibbs energy as
Eq.~\eqref{eq:aniG}, with the background $G_A(p,T)$ independent of the order
parameter. Criticality then occurs at $x=1/2$ with $T_c=\omega/2\kb$
[Ref.~\cite{Anisimov2018}, Eq.~(6)], so a liquid--liquid critical point requires
$\omega>0$, and $\omega=0$ is the singularity-free limit of their Fig.~3.

The present closure differs in one structural respect: the background depends on
the order parameter, through both $a(T,m)$ and $b(m)$. Equation~\eqref{eq:omegac}
is what replaces $T_c=\omega/2\kb$ once that is so, and the two agree where the
background is state-independent: with $b'=a'=0$ one has $f_{vm}=0$ and
$f_{mm}=T/(1-m^{2})$, so $\omega_c=2T$ at $m=0$, which is their Eq.~(6) in
reduced units.

A bilinear density--order coupling is already present in the non-degenerate
lattice model of Ref.~\cite{Caupin2021}, whose equations we cite by their numbers in
the appendix of its arXiv version. The mixed second derivative of the
Helmholtz energy of their Eq.~(15), evaluated at internal equilibrium, is
\begin{equation}
\left.\frac{\partial^{2}f}{\partial\rho\,\partial x}\right|_{\rm eq}
=-\rho\big[(\omega_1-\omega_2)+\omega\,(2x-1)\big],
\label{eq:caupinbilinear}
\end{equation}
which vanishes only on the line $x=x_\times$ of their Eq.~(20) and is therefore
non-zero at $x=1/2$ whenever $\omega_1\neq\omega_2$; their Eq.~(21) gives
$\rho(\partial x/\partial\rho)_T$, and $(\partial x/\partial\rho)_T$ is exactly
$-f_{\rho x}/f_{xx}$, the same elimination written in the conjugate direction,
which enters their critical conditions, their Eqs.~(24)--(25).

Where the two differ is in the channels available to the coupling. On a fixed
lattice with single occupancy, such as that of Ref.~\cite{Caupin2021}, every site has
the same size, so the translational term
$T[\rho\ln\rho+(1-\rho)\ln(1-\rho)]$ carries no dependence on $x$ and the whole
of $f_{\rho x}$ comes from the attractive mean field. Here $f_{vm}$ has two
channels, Eq.~\eqref{eq:rclsplit}, and the excluded-volume one has no lattice
counterpart. Over the $82$ states with $\omega_c<0$ that are global minima
within the liquid manifold on the converged mesh, it accounts for $8.5\%$ to
$99.98\%$ of $|f_{vm}|$, with median $91.1\%$, dominates in magnitude in $67$,
and carries a sign opposite to the attractive channel in $58$. Retaining only
the attractive channel in the Schur term, with $f_{mm}$ and $f_{vv}$ held at
their values --- a diagnostic of the weight of that term, not a model, since
setting $b'=0$ would also change $f_{mm}$ --- leaves $R_{\rm cl}>0$ in $58$ of
the $82$. The proportion is stable under refinement: on the mesh of $136$ the
same criterion gives $29$ of $42$, or $69.0\%$, against $70.7\%$ here. The
excluded-volume channel is therefore necessary for about seven in ten of the
states found, and the critical point of the fixed-parameter model of
Fig.~\ref{fig:neg}, at $73.0\%$, lies below the median of that
distribution rather than at its upper edge.

The lattice model of Ref.~\cite{Caupin2021} already admits such states through the
attractive route alone, although they were not reported there: with their energy
and entropy excesses equal to $2$ and $20$, $\omega_1=16.82$ and $\omega_2=2$, their
Eq.~(15) gives in mean field a liquid--liquid critical point at
$(\rho,T)=(0.914,0.935)$ for $\omega=0$, and the point persists for every $\omega$
examined down to $-2$ (our calculation). What has no lattice counterpart is the
excluded-volume route, which the closure studied here needs in seven of every ten
of the states found.

The comparison with Ref.~\cite{Anisimov2018} is exact at $m=0$, where the ideal part of
$R_{\rm cl}$ is $T$; for a prescribed state with $m\neq0$ the local ideal
reference is $2T/(1-m^{2})$ rather than $2T$, and the ratios quoted below should
be read with that in mind. Equation~\eqref{eq:omegac} reproduces the
regular-solution value wherever the closure contributes nothing to the
curvature, that is in the dilute limit:
$\omega_c/2T$ tends to $0.9985$, $0.9965$ and $0.9963$ at $v=300$ and $T=0.90\,\TLG$ for
$(\alpha,\gamma,\kappa)=(0.10,0.50,0.20)$, $(0.20,0.50,0.30)$ and $(0.20,0.70,0.40)$,
approaching unity from below --- at $v=10^{4}$ the three are $0.99996$, $0.99990$ and
$0.99989$ --- with no adjustment. These are
evaluated at $m=0$, where the ideal reference is unity; away from $m=0$ the local
ideal reference is $1/(1-m^{2})$ and the comparison must be made against that. At liquid
densities the closure renormalizes the threshold in either direction. The
comparison must be made against the local ideal reference, which is $1/(1-m^2)$
rather than unity, so the quantity to report is
$Q\equiv(1-m_c^{2})\,\omega_c/2T$, equal to unity exactly when the closure
contributes nothing. For the nine representative states of
Table~\ref{tab:states}, $Q$ spans $0.009$ to $1.742$ with median $0.287$, eight
of the nine lying below unity and one, the densest, well above it; quoting $\omega_c/2T$ in its place overstates the ratio wherever
$m_c$ is appreciable, by a factor of $1.6$ at the largest $|m_c|$ of the table.
Within this closure the critical value of $\omega$ is \emph{generated} as a
function of state from derivatives of the free energy, in closed form, rather
than supplied as a phenomenological parameter, in general a function of $T$ and $p$~\cite{Anisimov2018}.

As an ordering of
scales, the solutions reported lie at temperatures of the order of the reference
liquid--gas critical temperature, up to $1.15\,\TLG$; for water the ratio of the
liquid--liquid to the liquid--gas critical temperature is $198/647\simeq0.31$, with the first temperature from Ref.~\cite{NatPhys2025}
and the second from Ref.~\cite{IAPWS95}, lower than anything found here. Recent
experiments place that critical point at $210\pm8$~K, with a critical pressure of
$1002\pm151$~bar inferred from an extrapolated equation of
state~\cite{Science2026}, in a region metastable against crystallization. Metastability of a
liquid--liquid critical point is therefore familiar, though the competing phase
there is the crystal and here it is the vapor, and we expect the two to carry different
nucleation barriers.

\section{Discussion and conclusions}
\label{sec:conclusions}

Three results stand. First, the quantity that governs the internal mode after the
density mode has been relaxed is the Schur complement $R_{\rm cl}$, ideal part
included, and the non-ideality required at criticality is
$\omega_c=2R_{\rm cl}$. That relation is a structural identity of two-state free
energies of this form, the same elimination that enters the critical conditions of
Ref.~\cite{Caupin2021} [their Eqs.~(21) and (24)] and the Landau treatment of
Ref.~\cite{Anisimov2025} [their Appendix~A]; what the closure supplies is
$R_{\rm cl}(T,v,m)$ itself, which changes sign inside the liquid region, so that
the baseline passes from subcritical, needing a demixing-promoting non-ideality,
to overdriven, needing a mixing-promoting one. Below $T_+$, the temperature that bounds
the states requiring $\omega_c>0$, locally critical states persist in seven of
the nine sets examined; in those seven the upper envelope of $\omega_c$ changes
sign at $T_+$ and every root detected below it requires the opposite sign. In the
remaining two the critical solutions simply cease to be detected while $\omega_c$
is still positive, so $T_+$ is there the end of the solutions and not a sign
change. The two descriptions coincide wherever both exist, but they are not the
same statement. Second, the coupling that carries that sign change has two
channels, Eq.~\eqref{eq:rclsplit}: the attractive mean field and the
state-dependent excluded volume. Only the first has a counterpart on a fixed
lattice with single occupancy, where every site is the same size, and the second
is not redundant:
across the critical states that are global minima within the liquid manifold and
require $\omega_c<0$, retaining the attractive channel alone in the Schur term,
with $f_{mm}$ and $f_{vv}$ held at their values, leaves $R_{\rm cl}$ positive in
$58$ of the $82$.
Third, and this is the construction we would put first, a fully specified
fixed-parameter model whose non-ideality \emph{favors} mixing can still undergo a
liquid--liquid transition.
For two parameter sets, one in each sign sector, we exhibited the coexistence
line of a model with fixed $(\omega,\Delta H,\Delta S)$, terminating at the
predicted critical point with
the mean-field closing of the order-parameter gap. The threshold non-ideality is given
in closed form and, in the dilute limit, reduces to $Q\to1$, which at $m=0$ is
exactly the regular-solution condition. At a given critical state the closure fixes the combined field by stationarity,
while the enthalpy and entropy of interconversion, which are one-particle
properties, enter as inputs.

These results place a closure built from a pair potential inside the two-state
description of Refs.~\cite{Anisimov2018,Anisimov2025}, in which interconversion is
coupled to density, as it is in the minimal microscopic model of
Ref.~\cite{Caupin2021} and in maximum-valence treatments~\cite{Shumovskyi2022}; that
pair potentials can produce liquid--liquid transitions was established in
Ref.~\cite{Franzese2001}. What the closure adds, across the $34$ parameter sets
sampled, is the critical non-ideality as a closed-form function of state and the
excluded-volume route to a non-ideality that favors mixing.

Several features of the numerical work set how these results are to be read,
and Secs.~\ref{sec:closure}--\ref{sec:binodal} and Appendix~\ref{app:num} give the evidence for each. In the asymmetric sector we
report the presence counts as counts of parameter sets rather than numbers of critical states, because the
solutions form continuous branches in the order parameter and any count or ratio of
roots carries the measure induced by the mesh; the first three set counts are
unchanged under refinement and the two global ones settle at $136$ values of $m$,
where all five are also identical, set by set, on Chebyshev--Lobatto nodes. The
global-stability test uses two temperatures per parameter set, so it is a sampling
of parameter space and not a map, and $T_+$ is a boundary located over sampled
roots rather than the full locus $R_{\rm cl}=0$. Over the $34$ parameter sets
stability is established within the liquid manifold; against the vapor it was
examined for the two fixed-parameter models of Sec.~\ref{sec:binodal}, the one at
$\omega<0$ being lower in Gibbs energy than the vapor only within
$2.04\times10^{-4}$ of $T_c$, while for the one at $\omega>0$ no vapor state
competes. Four of the representative states of Table~\ref{tab:states} sit at
negative pressure. $T_+$ is quoted to three decimals, and differences below about
$0.03$ in $T/\TLG$ should not be read quantitatively. The classification of how
the envelope changes sign is a statement about the sampled solution set and not
about the branch: establishing termination as such would require continuation.
Root finding detects sign changes of the third directional derivative and could in
principle miss a tangential root of even multiplicity; a direct search for such
roots in the two sets that set the extremes of $T_+$ found none. The three-phase
point of Sec.~\ref{sec:binodal} is inferred rather than exhibited. Finally, we read ratios
and topology as exact within the closure, while absolute values carry
the bias of its second-virial input, which the higher-order treatment of
Ref.~\cite{SPYP2020} addresses.

Equation~\eqref{eq:rclsplit} holds for any closure of van der Waals form with
state-dependent $b(m)$ and $a(T,m)$ and the ideal mixing entropy of two states, so
for any such closure the sign of the non-ideality that criticality requires can be
computed rather than assumed.

\section*{Disclosure statement}

The author declares no competing financial interests or personal relationships
that could have appeared to influence the work reported here.

\section*{Funding}

This research received no specific grant from any funding agency.

\section*{Data availability}

The data and code that support the findings of this study are openly available.
A package of scripts rebuilds the theory from the integrals, validates it against
Ref.~\cite{SPD2001}, and generates every number and figure quoted here. It
contains the implementations described in Appendix~\ref{app:num},
the master table of roots from which every aggregate in Table~\ref{tab:audit} is
derived, and a README describing how the numerical claims map to the scripts that produce them. Notes on further numerical checks, beyond those of
Appendix~\ref{app:num}, accompany the package. It is archived at Ref.~\cite{Zenodo}, which should be cited together with this
paper by anyone using the data.

\section*{Declaration of generative AI use}

In accordance with the Taylor \& Francis policy on generative AI, the author
discloses that a large language model, Claude Opus 5.5 (Anthropic), was used to
speed up and assist with the analysis code, the symbolic eliminations, the figure
scripts, and the writing and editing of the text. The author defined the method and
the acceptance criteria, and verified all results against the replication package
deposited at Zenodo~\cite{Zenodo} and by the independent cross-checks reported in the
Appendices. The author alone is responsible for the content of this article; no
figure is an AI-generated image, and all figures are plots produced
deterministically by the scripts of that package.

\section*{ORCID}

\emph{Marcio Pazetti} \url{https://orcid.org/\ORCID}


\appendix
\numberwithin{equation}{section}

\section{Angular moments and their validation}
\label{app:moments}

The moments \eqref{eq:moments} are polynomials of degree three, four and five in
$\alpha$ under the measure of Ref.~\cite{SPD2001}. Computed symbolically and
compared with the printed expressions, they agree term by term, including
coefficients such as $-669\,524/2\,401\,245$ multiplying $\alpha^5\gamma^3$ in
$C_2$. The degree-five dependence follows whenever the core amplitude is linear in $\alpha$, as the
product $\gamma\alpha$ is; with $s=\gamma\alpha^{p}$ the degree is $3p+2$.

\section{Normalization chain}
\label{app:norm}

With the unnormalized measure implied by Ref.~\cite{SPD2001}, $\langle1\rangle=4$,
$C_0=4$ isotropically and $b=V_mC_0/8=V_m/2$. With the normalized state measure,
$\langle1\rangle=1$, $C_0=1$ and $b=V_mC_0/2=V_m/2$. The hard-core second virial
coefficient is $2\pi r_0^3/3$ in both.

\section{Numerical procedure}
\label{app:num}

This appendix collects the numerical procedures and the convergence tests
behind the results of Secs.~\ref{sec:crit}--\ref{sec:compare}.

\subsection{Parameter grids}
\label{app:grids}

Four grids appear in this work and they are not the same;
Table~\ref{tab:grids} identifies them.
The first two are sampled subsets of a coarse lattice in $(\alpha,\gamma,\kappa)$
rather than full Cartesian products, which is why their sizes are not products of
the numbers of sampled values; the third is their union, used only for the
sensitivity of the results to the cut on $A_{\min}$; and the fourth is the
symmetric sector, a different construction that carries no mesh in the order
parameter.

\begin{table}[htbp]
\centering
\caption{%
The parameter grids used. Counting units are separated, because
conflating them makes the table fail to factorize:
$N_{\rm sets}$ counts distinct $(\alpha,\gamma,\kappa)$ triples, $N_T$ is the
number of temperatures at which each was searched, and the evaluation counts are
the products. The grid of Table~\ref{tab:audit} is the admissible part of the
main grid, its $34$ triples with $A_{\min}>0.3$; the nine states of
Table~\ref{tab:states} are eight from the main grid and one,
$(0.15,1.40,0.30)$, from the gap-filling grid. The cut $A_{\min}>0.3$ acts on the triple alone, since
$A_{\min}=1-\gamma\alpha-\kappa$ carries no temperature, so it removes whole
triples and both evaluation columns stay products of $N_{\rm sets}$ and $N_T$. In the symmetric grid the cut is
inactive: with $\kappa=0$ the smallest $A_{\min}$ on the grid is $0.72$. The set
$(\alpha,\gamma,\kappa)=(0.15,1.40,0.30)$ of Table~\ref{tab:states} belongs to
the gap-filling grid, not to the main one. Every entry is generated from the
archived tables.
}
\label{tab:grids}
{\small\setlength{\tabcolsep}{5pt}
\begin{tabular*}{\textwidth}{@{\extracolsep{\fill}}lp{0.30\linewidth}ccccc}
\toprule
grid & $\alpha$, $\gamma$, $\kappa$ sampled & $N_{\rm sets}$ &
with & $N_T$ & evals. & evals.\\
 & & raw & $A_{\min}>0.3$ & & raw & admissible\\
\midrule
main & \raggedright sampled from $\alpha\in\{0.10,\dots,0.50\}$,
       $\gamma\in\{0.35,\dots,2.50\}$, $\kappa\in\{0.02,\dots,0.40\}$
     & 45 & 34 & 2 & 90 & 68\\[2pt]
gap-filling & \raggedright $\alpha\in\{0.15,0.25,0.35,0.45\}$,
              $\gamma\in\{1.00,1.40,1.80\}$, $\kappa\in\{0.10,0.20,0.30\}$;
              contains $(0.15,1.40,0.30)$ & 21 & 21 & 2 & 42 & 42\\[2pt]
union & \raggedright the two above together, used for the sensitivity to the cut
      & 66 & 55 & --- & --- & ---\\[2pt]
symmetric & \raggedright $\alpha$: $11$ values, $0.02$--$0.20$; $\gamma$: $5$ values,
            $0.35$--$1.40$; $\kappa=0$ & 55 & 55 & 5 & 275 & 275\\
\bottomrule
\end{tabular*}}
\end{table}

The cut $A_{\min}>0.3$ of Sec.~\ref{sec:sym2} was varied over six other values.
Absolute counts are the wrong statistic, since tightening the cut also removes
admissible sets; the informative quantity is the fraction of admissible sets that
admit solutions. That fraction stays between $0.649$ and $0.833$ as the cut is tightened: it is
$0.661$, $0.661$, $0.649$ and $0.655$ for cuts of $0.15$ to $0.30$ and rises to
$0.729$, $0.733$ and $0.833$ at $0.35$, $0.40$ and $0.45$, over $62$ down to $36$
admissible sets, all at the two temperatures of the grid; on the common subset
surviving every cut, $A_{\min}>0.45$, it is $30$ of $36$.

\subsection{Meshes and root search}
\label{app:mesh}

Meshes are reported per result, not globally.
Different parts of the calculation use different discretizations, and quoting a
single ``production mesh'' would misdescribe all of them. The search of
the asymmetric sector uses $\xi=400$ and, for the counts of
Table~\ref{tab:audit}, $m=136$, checked against $m=271$. In both searches the mesh in $\xi$ runs
from $\xi_{\min}=10^{-4}$ to $(1-10^{-7})\,\xi_{\max}(T,m)$, where
$\xi_{\max}$ is the end of the mechanically stable liquid branch at that
$(T,m)$, located by bisection on $f_{vv}=0$. Three features of that mesh are not
free choices: two are forced by solutions that would otherwise be
missed, and the third by nodes that would otherwise fall outside the branch. The upper end is \emph{adaptive}, so the number of nodes alone does not
determine the mesh: at the first row of Table~\ref{tab:states}
$\xi_{\max}=0.475$, and the same $30$ nodes carried to a fixed upper end of $20$
would place only $21$ of them inside the branch. The end itself must be the
spinodal and not a fixed fraction of it: the roots that fix $T_+=0.745$ and
$0.583$ lie within $0.02\%$ of the spinodal, at $\xi/\xi_{\max}\simeq0.9998$,
where $f_{vv}$ is of order $10^{-3}$, so a ceiling at $0.999\,\xi_{\max}$ would
exclude them. And the mesh is \emph{densified over the last $2\%$ of the
branch}: to the geometric nodes it adds as many again, equally spaced from
$0.98\,\xi_{\max}$ to the upper end, so that $\xi=400$ stands for $400+400$
values, because there the critical family develops pairs of simple roots
separated by $\sim10^{-5}$ in $\xi/\xi_{\max}$: a dip narrower than one interval
of the geometric mesh produces neither a sign change nor a local minimum, and is
invisible to both tests. Pairs that still fall inside one interval are recovered
by refining the local minima of $|D^3f|$ that produce no sign change, on
geometric sub-meshes of $120$ values, up to five levels deep. For those
two parameter sets the mesh in $m$ is refined locally as well, with a step below
the width of the window; the two values depend on those
solutions, by $0.025$ and $0.005$, and the other seven are insensitive to the treatment. Table~\ref{tab:states} is built
on $m=60$ and $\xi=30$ followed by $28$ bisection steps, so that the $\xi$ mesh
only brackets and the bisection measures, while the $m$ mesh quantizes: the
reported $m_c$ is a mesh node, which is why the selection rule needs the tie-break stated below. The coexistence constructions use
$M=441$ values of $m$ for the model at $\omega>0$ and $M=1201$ for the model at
$\omega<0$; the latter resolves all seventeen temperatures of
Fig.~\ref{fig:neg}, whereas $M=601$ resolves only fifteen, failing at the two
largest, and $M=1201$ and $M=2401$ agree with each other to $6\times10^{-13}$ in
the coexistence pressure.

Recomputing all nine rows of Table~\ref{tab:states} at $400$ values of $\xi$
leaves $m_c$ unchanged exactly, since it is a node of the $m$ mesh, and moves no
other entry by more than $1.5\times10^{-8}$, nearly five orders below the last printed digit.
Where a set has more than one root at the temperature quoted, the table lists,
among the roots requiring $\omega_c>0$, the one of smallest $|m|$. That rule needs
a tie-break, and the tie is not exceptional: roots can share the same $|m|$ in two
distinct ways. The mesh in $m$ is symmetric about the origin, so a mirror pair at
$m=\mp 0.0158$ can occur, and does; independently, two roots can sit at the same
$m$ with different $v$. Ties of either kind occur in four of the nine sets at the
temperatures listed, and the tied roots are distinct critical states, differing in
$\ueff$ by up to $0.42$ and in $v/b(m_c)$ by up to $0.23$, both extremes being
carried by the same-$m$ pair. Among tied roots we list the one of smallest $m$, and
among roots at equal $m$ the one of smallest $v$; the producer script archived in
the replication package applies that key explicitly.

At fixed temperature the search uses $68$ values of $m$ and $400$ of $\xi$, with
sign changes of the third directional derivative refined by bisection. The number
of $\xi$ values matters more than it appears to: on a plain geometric mesh, at $18$
and $36$ the bracketing misses sign changes rather than merely displacing them. At
$(\alpha,\gamma,\kappa)=(0.10,0.50,0.20)$ and $T=0.855\,\TLG$ such a mesh of $36$
misses roots that $120$ finds: $120$ returns nine, five of them positive, and
$120$, $400$ and $1200$ agree; with the densification and the recovery of close
pairs described above, all nine are found already at $18$. A coarse
scan of fourteen temperatures between $0.35$ and $0.97\,\TLG$ brackets the lower
edge, and bisection then locates it to $10^{-4}\,\TLG$; the values of
Table~\ref{tab:interval} are located instead by descending from above in steps of
$0.005\,\TLG$ with the window in $m$ tracked, and bisecting to $10^{-4}$. The upper
edge is bracketed on a scan from $0.95$ to $1.20\,\TLG$ in steps of $0.01$ and
bisected to $0.001$. The reference $\TLG$ is obtained by bisection on the closure
condition, within a bracket whose two endpoints are finite and of opposite sign,
with the residual validated.

\subsection{Convergence}
\label{app:conv}

The bisection tolerance bounds only the outer loop: $10^{-4}\,\TLG$ for the
lower edges of Table~\ref{tab:interval}, for those compared below at
$68\times400$ and for the sequence of uniform meshes in $m$ of
Sec.~\ref{sec:locus}, and $0.001$ for the upper edge. The inner test --- does a
solution exist at this temperature --- rests on a finite mesh in $m$ and $\xi$ and
on sign changes of $D^3f[\mathbf{e},\mathbf{e},\mathbf{e}]$, and the two mesh
directions do not behave alike. The $\xi$ direction is the one along which roots
are bracketed, and a coarse $\xi$ mesh does not merely displace a root: it can fail
to bracket a sign change at all and drop the root from the count, as the case of
Appendix~\ref{app:mesh} shows. Convergence is therefore studied in each mesh
direction separately: in $\xi$ at $120$, $400$ and $1200$ values, at $68$ values of
$m$, where all three agree root by root, and in $m$ at $40$, $68$, $136$ and $271$,
where the first three presence counts of Table~\ref{tab:audit} do not move at all
and the two global counts settle at $136$, unchanged at $271$.

The two global counts are $10$ and $8$ at $40$ values of $m$, $12$ and $9$ at
$68$, and $12$ and $10$ at both $136$ and $271$. Being the global minimum is a property of a window in $m$ along the critical family, and a
coarse mesh can step over such a window: at $(0.30,0.50,0.20)$ the globally preferred roots
lie near $m=0.834$ and $m=0.778$, which the meshes of $40$, $136$ and $271$ values sample
and that of $68$ straddles, missing by $4.3\times10^{-4}$ in $g$. Because the meshes of
$136$ and $271$ are nested, every root found on the first is found on the second, and the
two global counts are lower bounds that can only rise under such refinement; that they do
not rise, set by set, is the evidence we have for convergence.

The boundary $T_+$ was tested in the $\xi$ direction as well: on a plain
geometric mesh at $68$ values of $m$, refining $\xi$ from $18$ to $400$ values moved
$T_+$ in two of the seven sets in which the envelope changes sign, by $0.112$ and
$0.025$, which is why coarse meshes are not used here. For
$(\alpha,\gamma,\kappa)=(0.20,0.50,0.30)$ and $(0.20,0.50,0.40)$ the mesh
$68\times400$ gives $0.694$ and $0.596$, the latter with roots also on two isolated
intervals below it, from $0.589$ to $0.593$ and near $0.584$, in a scan in steps of
$10^{-4}$; the value of Table~\ref{tab:interval} for the second set, $0.583$, is
obtained by refining the mesh in $m$ locally, as for $(0.10,0.50,0.20)$
(Sec.~\ref{sec:locus}).

The number of roots found is not reported because it measures the mesh: the
solutions of Eq.~\eqref{eq:crit} form continuous branches in $m$, each mesh line
samples them, and refining the mesh in $m$ from $40$ to $68$ values multiplies the
count by $1.681$, against a mesh ratio of $1.700$. Nor do proportions repair this.
A ratio such as (roots with $\omega_c>0$)/(all roots) is a measure induced by the
sampling, and changing the \emph{spacing} rather than the number of nodes changes
it: replacing the uniform mesh by Chebyshev--Lobatto nodes, at the same $40$ nodes,
moves that ratio by $9.1$ percentage points over the $34$ parameter sets, about
eleven times the shift of $0.8$ points produced by refining the uniform mesh from
$40$ to $68$ values, and at $136$ nodes it still moves it by $9.6$. The set counts
are robust to the same change: at $136$ nodes all five are identical, set by set,
on the two meshes; at $40$ nodes, where the global counts have not yet converged
and move from $10$ and $8$ to $10$ and $10$, the second count also falls by one,
from $21$ to $20$.

Root finding detects sign changes of the third directional derivative and could
in principle miss a tangential root of even multiplicity. We searched for such
roots directly, by looking for local minima of $|D^3f[\mathbf{e},\mathbf{e},\mathbf{e}]|$
that descend below $10^{-3}$ of the typical scale without changing sign, over ten
scans at mesh $68\times200$, with the plain geometric mesh in $\xi$ (no
densification and no recovery of pairs), at five equally spaced temperatures from
$0.002\,\TLG$ above $T_+$ to $0.002\,\TLG$ below the upper edge in each
of the two sets that set the extremes of $T_+$ in Table~\ref{tab:interval}, the
typical scale being the median of $|D^3f|$ along each line. Across $504$ ordinary
sign changes no such candidate appeared; the deepest local minimum without a sign
change stays at $3\times10^{-3}$ of that scale.

\subsection{Mechanical roots, residuals and the global test}
\label{app:global}

The global-stability filter of Sec.~\ref{sec:binodal} is applied within the same
unrestricted search, to every root it finds, at every parameter set with
$A_{\min}>0.3$ and at both temperatures, $0.80$ and $0.92\,\TLG$. Every root admits
the global test: the liquid branch through $(m_c,v_c)$ is guaranteed by the implicit
function theorem, since the search retains only points with $f_{vv}>0$ and sets
$p_c=-f_v$ there. The liquid volume is not searched for. Since $p=T/(v-b)-a/v^{2}$, the
condition $p(v,m)=p_c$ is the cubic
\begin{equation}
p_c\,v^{3}-(p_c b+T)\,v^{2}+a\,v-a\,b=0 ,
\label{eq:mechcubic}
\end{equation}
with $a=a(T,m)$ and $b=b(m)$, whose roots are enumerated in closed form; those
with $v>b$, $f_{vv}>0$ and pressure residual below $10^{-8}$ are the mechanically
stable ones, and the smallest is the liquid.  Three residuals were recorded for every root of the grid,
without sampling, before any minimization: $r_p=|p(v_c,m_c)-p_c|$,
$r_1=|g'(m_c)|$ and $r_2=|g''(m_c)|$, with $g'$ and $g''$ from
Eq.~\eqref{eq:gderivs}. Over the whole population at all four resolutions of the order-parameter mesh
--- $4161$ roots at $40$, $68$, $136$ and $271$ values of $m$ --- $r_p=0$ and
$r_2=0$ exactly and $r_1<7.2\times10^{-15}$, as Eq.~\eqref{eq:gderivs}
requires; the cubic reproduces $v_c$ for every root within the tolerance
$10^{-8}$ imposed on the search, the worst relative deviation being
$7.2\times10^{-9}$; and no root fails to admit the constrained Gibbs energy at
its own order parameter. Where two mechanically stable roots exist at the same
pressure, the smaller is the liquid one, and it is the one the search had
selected in every case.

In the symmetric sector the zeros of $\ueff$ on the liquid branch at $m=0$ are counted by a
Sturm sequence on the numerator of $f_{vv}f_{mmmm}-3f_{vmm}^{2}$, a quartic in $v$, over
$b_0<v<v_{\rm spin}$, so that nothing is discretized, with $\hat S$
summed by the recursion $t_{n+1}=t_n\,z(2n-1)/[(n+1)(2n+1)]$ taken to $n=300$; at
every argument used, the terms beyond $n=32$ are below $10^{-15}\max(1,|\hat S|)$. The global test at a candidate fixes $p=-f_v$ and $\omega=\omega_c$ at the root,
with the field required by $\Phi_m=0$, and then locates the global minimum of
$g$ over the liquid manifold as follows. A uniform mesh of $401$ values of $m$ on $|m|\le0.9985$, supplemented by geometric tails reaching $1-|m|=10^{-13}$ and by the
endpoints $m=\pm1$ themselves, is used to bracket the basins: every interior local minimum it brackets is then refined by a ternary
search to a width below $10^{-12}$, and the two endpoints of the interval on
which $g$ is defined are evaluated as well, so that minima approaching
$m\to\pm1$ are not missed. The value at the candidate is compared with the
smallest of all those values, nodal and refined. The tails are not ornamental: for one root of
the grid of $68$ values of $m$, $(\alpha,\gamma,\kappa)=(0.30,0.70,0.40)$ at
$0.92\,\TLG$, the
competing state is the nearly fully ordered liquid, which first undercuts the
root at $1+m=3.6\times10^{-4}$ and reaches its minimum at $m=-1$, a margin of
$1.2\times10^{-2}$ that a uniform mesh of spacing $5\times10^{-3}$ cannot
resolve. On the mesh of $68$ values of $m$ the tails lower the reported minimum for
$387$ of the $548$ roots and change one classification; on the mesh of $136$ values
they lower it for $786$ of the $1101$ roots and change two, in neither case altering
a count of Table~\ref{tab:audit}. The mesh therefore locates, and the refinement
decides; the procedure is not a comparison of mesh values. It remains an
enumeration of the basins the mesh resolves, and a minimum narrower than the mesh
spacing between two sampled points could still escape it.

\subsection{Coexistence}
\label{app:coex}

Coexistence is located from $\Delta g=0$.
For each temperature the two liquid minima of Eq.~\eqref{eq:gibbs} are tracked
from the critical point by continuity in $m$, and the coexistence pressure is the
root of $\Delta g(p)=g_R(p)-g_L(p)$, bisected until the bracket in $p$ is narrower
than $10^{-11}$ in relative terms. Each minimum is refined inside a bracket taken
from the geometry of the function --- the two mesh nodes flanking a local minimum
--- and never inside a window of prescribed width, since a fixed window returns
its own edge whenever the minimum falls outside it, and an edge is not a
stationary point. Every accepted solution is required to have both minima
interior to their brackets, both with positive curvature, the same two branches
maintained throughout the bracketing, and $|\Delta g|$ at the located pressure
below $10^{-9}$; in practice the residual is at the level of machine precision,
below $3\times10^{-14}$ in all reported points. A reseed that would require either
branch to move further than $0.05$ in $m$ is refused and reported rather than
followed, which is what keeps the liquid--liquid line from continuing silently
into the liquid--dense one. The critical root of each fixed-parameter model is
fixed by an intrinsic rule. For the model at $\omega>0$ it is the one of smallest
$v/b(m)$ --- the innermost critical point on the liquid branch; two critical roots
exist at every $m$ in the relevant range, so the choice is a real one, and at the
critical point of the model only that one has $\ueff>0$. For the model at
$\omega<0$ the global test does not decide: at $m=-0.93$ and $0.92\,\TLG$ three
critical roots, at $v/b(m)=1.196$, $1.548$ and $1.601$, are each the global minimum
of Eq.~\eqref{eq:gibbs} at their own pressure, and the model is the one of largest
$\omega_c$, which is also the largest $\omega_c$ among all critical roots with
$\omega_c<0$ of that parameter set and temperature on every mesh used.

At the critical point of the model at $\omega>0$ every residual sits at machine
precision: $g'(m_c)=-2\times10^{-17}$, $g''(m_c)=0$ to the last bit, $p+f_v=0$,
$D^3f[\mathbf{e},\mathbf{e},\mathbf{e}]=6\times10^{-16}$, and the determinant of
$H_\omega$ is $-3\times10^{-17}$. Because the critical point is solved for that
$\omega$, the identity $\omega=2R_{\rm cl}$ holds to machine precision by
construction, as Appendix~\ref{app:global} confirms it does over the whole
population of roots. The minima of Table~\ref{tab:coex} are located by solving
$\mathrm{d}g/\mathrm{d}m=0$, with the field entering as $+\Delta H/2$, rather than
by a minimizer: at coexistence the wells are flat to order $10^{-13}$ in $g$, which
is below the resolution of any comparison of function values. Following the
liquid--liquid pair of the model at $\omega>0$ beyond
$1-T/T_c\simeq1.16\times10^{-3}$, where it is metastable against the dense phase,
requires selecting the pair by proximity in $m$ to the critical point rather than
by taking the two lowest minima, which beyond that boundary would return the dense
phase instead; the producer archived in the replication package applies that rule
and labels the points accordingly.

\subsection{Comparison with the gas branch}
\label{app:gas}

The statement in Sec.~\ref{sec:binodal} that the liquid--liquid coexistence of
the model at $\omega<0$ is lower in Gibbs energy than the vapor only within
$2.04\times10^{-4}$ of $T_c$ is obtained as follows. At a temperature and the
coexistence pressure located as above, Eq.~\eqref{eq:mechcubic} is solved for the
mechanical roots at each of $801$ values of $m$, and the \emph{largest} root
satisfying $v>b(m)$, $f_{vv}>0$ and a pressure residual below $10^{-8}$ is taken
as the gas state, in contrast with the smallest, which is the liquid. The same
functional as in Eq.~\eqref{eq:gibbs} is then evaluated on that branch, including
the field and the non-ideality, and minimized over $m$ by the same procedure used
for the liquid. The quantity compared is
$\Delta g_{\rm gas-liq}=\min_m g_{\rm gas}-\min_m g_{\rm liq}$, positive when the
liquid is the stable phase. Along the coexistence line of Fig.~\ref{fig:neg} it
takes the values $+7.9\times10^{-4}$, $-2.6\times10^{-3}$, $-2.3\times10^{-2}$
and $-9.2\times10^{-2}$ at $1-T/T_c$ equal to $1.0\times10^{-4}$,
$5.4\times10^{-4}$, $2.9\times10^{-3}$ and $9.1\times10^{-3}$. The crossing is
then located by bisection in $\log(1-T/T_c)$, the liquid--liquid coexistence
pressure being located afresh at every temperature tested, from the discontinuity of $\arg\min_m g$, rather than interpolated,
to a relative bracket width of $2.6\times10^{-5}$; it lies at
$1-T/T_c=2.042\times10^{-4}$, where the coexistence pressure is $0.153739$. For the model at
$\omega>0$ the same enumeration returns a distinct gas root only for
$m\lesssim-0.92$, and $\min_m g_{\rm gas}-\min_m g_{\rm liq}$ is at least $0.085$
along the whole line, its smallest value at the critical point, so no gas state
competes.

The gas-branch comparison at every point of the coexistence line of
Fig.~\ref{fig:neg}, from which the boundary quoted above is read off, is
tabulated in the Supplemental Material.

\subsection{The mean-field exponent}
\label{app:exp}

Fitting the order-parameter gap of Sec.~\ref{sec:binodal} over
$1-T/T_c\in[1.2\times10^{-4},10^{-3}]$ gives $\Delta m\sim(1-T/T_c)^{\beta}$ with
values \emph{consistent with the mean-field value $\beta=1/2$} in both models. We
report the exponent qualitatively rather than as a number with three decimals,
and the reason is worth stating, because it is not a limitation of the
arithmetic. The coexistence pressure and the two minima are converged: over the
meshes $M=441$, $601$, $1201$ the fitted exponent moves by $5\times10^{-6}$,
the coexistence pressure by $6\times10^{-13}$, and the largest residual
$|\Delta g|$ between the two minima is $3\times10^{-14}$. What
does move, by a few times $10^{-3}$, is the value returned by different fitting
windows over the same converged data --- two orders of magnitude above the
numerical resolution. We read that spread as the curvature of the power law across a
window narrower than two decades in $1-T/T_c$, not as an uncertainty in the
exponent, and quoting a third decimal would attribute to the measurement a
precision that belongs to the choice of window. A mean-field functional must give
$\beta=1/2$ at an ordinary critical point; the fit is retained in the archived
scripts as a diagnostic that the coexistence construction terminates at the
critical point the search predicted, with no target value anywhere in the
pipeline~\cite{Zenodo}.

\subsection{Computational tools}
\label{app:tools}

The numerical work was carried out in independent implementations across two
systems. The production sweeps, the root searches and the coexistence constructions
were written in Python (NumPy, SymPy, Matplotlib); the exact-arithmetic
checks were carried out in Wolfram Mathematica 15.0.1 (Windows, 64-bit). In the
Mathematica implementation, and only there, the numerical
work was done at \texttt{WorkingPrecision} $=40$, except the compiled evaluations, the global test among them, which run at machine precision; the Python
production runs use double precision throughout, and the convergence claims made
for them are mesh-refinement claims, not precision claims. No code is shared
between the implementations.

\end{document}


\received{Compiled November 10, 2026}

\title{Supplemental Material for ``Liquid--liquid criticality in a fluid whose
two structures favor mixing''}
\author{
\name{Marcio Pazetti}
\affil{Colegiado de Engenharia de Produ\c{c}\~ao, Universidade Federal do Vale do S\~ao Francisco,
Av.\ Ant\^onio Carlos Magalh\~aes 510, Country Club, Juazeiro, BA 48902-300, Brazil}
}
\maketitle

This document collects the point-by-point enumerations behind two statements of
the main text. Every entry is regenerated by the replication package cited
there; nothing here is needed to follow the argument.

\section{Set-by-set listing behind the presence counts}

Table~1 of the main text reports the column totals only. The listing below gives
the presence of each kind of critical state in each of the thirty-four parameter
sets, and is the source of those totals.

{\small\setlength{\tabcolsep}{5pt}
\begin{longtable}{@{\extracolsep{\fill}}ccccccccc}
\caption{Presence of each kind of critical state in each parameter set: $34$ sets with $A_{\min}>0.3$, two temperatures each at $0.80$ and
$0.92\,\TLG$, order-parameter meshes of $136$ and $271$ values, which agree, and
$400$ values of $\xi$.
A tick means that at least one root of the stated kind was found. The columns do
not partition the sets: twelve carry roots of both signs. The first three column
totals are unchanged at $40$, $68$, $136$ and $271$ values of $m$; the two global
totals are $10$ and $8$ at $40$, $12$ and $9$ at $68$, and $12$ and $10$ at both
$136$ and $271$, as Section~4 and Appendix~C.3 of the main text discuss. This table is the
source of the aggregates quoted there for $136$ and $271$ values of $m$; absolute numbers of roots are not reported, for the
reason given in Section~4.}\label{tab:auditS}\\
\toprule
$\alpha$ & $\gamma$ & $\kappa$ & $A_{\min}$ & any root & $\omega_c>0$ &
$\omega_c<0$ & global, $\omega_c>0$ & global, $\omega_c<0$\\
\midrule
\endfirsthead
\multicolumn{9}{c}{\textit{Table \ref{tab:auditS}, continued}}\\
\toprule
$\alpha$ & $\gamma$ & $\kappa$ & $A_{\min}$ & any root & $\omega_c>0$ &
$\omega_c<0$ & global, $\omega_c>0$ & global, $\omega_c<0$\\
\midrule
\endhead
\bottomrule
\endlastfoot
0.10 & 0.35 & 0.20 & 0.765 & $\checkmark$ & $\checkmark$ & $\checkmark$ & $\checkmark$ & $\checkmark$\\
0.10 & 0.35 & 0.30 & 0.665 & $\checkmark$ & $\checkmark$ & $\checkmark$ & $\checkmark$ & $\checkmark$\\
0.10 & 0.35 & 0.40 & 0.565 & $\checkmark$ & $\checkmark$ & --- & $\checkmark$ & ---\\
0.10 & 0.50 & 0.20 & 0.750 & $\checkmark$ & $\checkmark$ & $\checkmark$ & $\checkmark$ & $\checkmark$\\
0.10 & 0.50 & 0.30 & 0.650 & $\checkmark$ & $\checkmark$ & $\checkmark$ & $\checkmark$ & $\checkmark$\\
0.10 & 0.50 & 0.40 & 0.550 & $\checkmark$ & $\checkmark$ & --- & $\checkmark$ & ---\\
0.10 & 0.70 & 0.20 & 0.730 & $\checkmark$ & $\checkmark$ & $\checkmark$ & $\checkmark$ & $\checkmark$\\
0.10 & 0.70 & 0.30 & 0.630 & $\checkmark$ & $\checkmark$ & --- & $\checkmark$ & ---\\
0.10 & 0.70 & 0.40 & 0.530 & $\checkmark$ & $\checkmark$ & --- & $\checkmark$ & ---\\
0.20 & 0.35 & 0.20 & 0.730 & $\checkmark$ & $\checkmark$ & $\checkmark$ & --- & ---\\
0.20 & 0.35 & 0.30 & 0.630 & $\checkmark$ & $\checkmark$ & --- & --- & ---\\
0.20 & 0.35 & 0.40 & 0.530 & $\checkmark$ & $\checkmark$ & --- & --- & ---\\
0.20 & 0.50 & 0.20 & 0.700 & $\checkmark$ & $\checkmark$ & $\checkmark$ & --- & ---\\
0.20 & 0.50 & 0.30 & 0.600 & $\checkmark$ & $\checkmark$ & --- & --- & ---\\
0.20 & 0.50 & 0.40 & 0.500 & $\checkmark$ & $\checkmark$ & --- & $\checkmark$ & ---\\
0.20 & 0.70 & 0.20 & 0.660 & $\checkmark$ & --- & $\checkmark$ & --- & $\checkmark$\\
0.20 & 0.70 & 0.30 & 0.560 & $\checkmark$ & $\checkmark$ & $\checkmark$ & $\checkmark$ & ---\\
0.20 & 0.70 & 0.40 & 0.460 & $\checkmark$ & $\checkmark$ & $\checkmark$ & $\checkmark$ & ---\\
0.30 & 0.35 & 0.20 & 0.695 & $\checkmark$ & --- & $\checkmark$ & --- & ---\\
0.30 & 0.35 & 0.30 & 0.595 & $\checkmark$ & --- & $\checkmark$ & --- & ---\\
0.30 & 0.35 & 0.40 & 0.495 & $\checkmark$ & $\checkmark$ & --- & --- & ---\\
0.30 & 0.50 & 0.20 & 0.650 & $\checkmark$ & --- & $\checkmark$ & --- & $\checkmark$\\
0.30 & 0.50 & 0.30 & 0.550 & $\checkmark$ & --- & $\checkmark$ & --- & ---\\
0.30 & 0.50 & 0.40 & 0.450 & $\checkmark$ & $\checkmark$ & $\checkmark$ & --- & ---\\
0.30 & 0.70 & 0.20 & 0.590 & $\checkmark$ & --- & $\checkmark$ & --- & $\checkmark$\\
0.30 & 0.70 & 0.30 & 0.490 & $\checkmark$ & --- & $\checkmark$ & --- & $\checkmark$\\
0.30 & 0.70 & 0.40 & 0.390 & $\checkmark$ & --- & $\checkmark$ & --- & $\checkmark$\\
0.30 & 1.50 & 0.02 & 0.530 & --- & --- & --- & --- & ---\\
0.30 & 1.50 & 0.06 & 0.490 & --- & --- & --- & --- & ---\\
0.30 & 1.50 & 0.12 & 0.430 & --- & --- & --- & --- & ---\\
0.30 & 2.00 & 0.02 & 0.380 & $\checkmark$ & $\checkmark$ & $\checkmark$ & --- & ---\\
0.30 & 2.00 & 0.06 & 0.340 & $\checkmark$ & $\checkmark$ & $\checkmark$ & --- & ---\\
0.40 & 1.50 & 0.02 & 0.380 & --- & --- & --- & --- & ---\\
0.40 & 1.50 & 0.06 & 0.340 & --- & --- & --- & --- & ---\\
\midrule
\multicolumn{4}{r}{\textbf{sets}} & \textbf{29} & \textbf{21} & \textbf{20} &
\textbf{12} & \textbf{10}\\
\end{longtable}}

\section{Gas branch along the liquid--liquid coexistence line}

The main text reports that the liquid--liquid coexistence of the model at
$\omega=-1.05296$ is lower in Gibbs energy than the vapor only within
$1-T/T_c<2.042\times10^{-4}$, and is metastable against vaporization outside
that window. The table below gives the comparison at every point of the
coexistence line, so that the boundary can be read off directly.

\begin{table}[htbp]
\centering
\caption{Liquid--liquid coexistence of the model at $\omega=-1.05296$ compared
with the vapor at the same temperature and coexistence pressure, at every point
of Fig.~2 of the main text. Here $g_{\rm liq}$ and $g_{\rm gas}$ are the minima over
$m$ of the constrained Gibbs energy on the liquid and gas branches, in units of
$\mu_0$; the liquid is the stable phase where the difference is positive. Three of the seventeen points lie inside the window
$1-T/T_c<2.042\times10^{-4}$; the rest describe the metastable extension. The
mixing entropy is normalized to vanish at $m=0$, so the values of $g$
exceed those of the convention that sets it to zero at $m=\pm1$ by $T\ln 2$;
differences at fixed $T$ are unaffected.}
\label{tab:gas}
{\small\setlength{\tabcolsep}{5pt}
\begin{tabular*}{\textwidth}{@{\extracolsep{\fill}}ccccccl}
\toprule
$1-T/T_c$ & $T$ & $p_{\rm coex}$ & $g_{\rm liq}$ & $g_{\rm gas}$ &
$g_{\rm gas}-g_{\rm liq}$ & status\\
\midrule
$1.00\times10^{-4}$ & 0.43451 & 0.15429 & $-1.124687$ & $-1.123898$ & $+7.89\times10^{-4}$ & stable\\
$1.33\times10^{-4}$ & 0.43449 & 0.15412 & $-1.124802$ & $-1.124259$ & $+5.43\times10^{-4}$ & stable\\
$1.76\times10^{-4}$ & 0.43447 & 0.15389 & $-1.124955$ & $-1.124738$ & $+2.17\times10^{-4}$ & stable\\
$2.33\times10^{-4}$ & 0.43445 & 0.15359 & $-1.125157$ & $-1.125374$ & $-2.18\times10^{-4}$ & metastable\\
$3.08\times10^{-4}$ & 0.43442 & 0.15318 & $-1.125424$ & $-1.126220$ & $-7.96\times10^{-4}$ & metastable\\
$4.09\times10^{-4}$ & 0.43437 & 0.15265 & $-1.125779$ & $-1.127346$ & $-1.57\times10^{-3}$ & metastable\\
$5.42\times10^{-4}$ & 0.43431 & 0.15194 & $-1.126250$ & $-1.128847$ & $-2.60\times10^{-3}$ & metastable\\
$7.18\times10^{-4}$ & 0.43424 & 0.15101 & $-1.126873$ & $-1.130850$ & $-3.98\times10^{-3}$ & metastable\\
$9.52\times10^{-4}$ & 0.43414 & 0.14977 & $-1.127700$ & $-1.133532$ & $-5.83\times10^{-3}$ & metastable\\
$1.26\times10^{-3}$ & 0.43400 & 0.14812 & $-1.128796$ & $-1.137133$ & $-8.34\times10^{-3}$ & metastable\\
$1.67\times10^{-3}$ & 0.43382 & 0.14594 & $-1.130248$ & $-1.141988$ & $-1.17\times10^{-2}$ & metastable\\
$2.21\times10^{-3}$ & 0.43359 & 0.14306 & $-1.132175$ & $-1.148573$ & $-1.64\times10^{-2}$ & metastable\\
$2.94\times10^{-3}$ & 0.43327 & 0.13924 & $-1.134730$ & $-1.157573$ & $-2.28\times10^{-2}$ & metastable\\
$3.89\times10^{-3}$ & 0.43286 & 0.13418 & $-1.138119$ & $-1.170001$ & $-3.19\times10^{-2}$ & metastable\\
$5.16\times10^{-3}$ & 0.43231 & 0.12749 & $-1.142617$ & $-1.187408$ & $-4.48\times10^{-2}$ & metastable\\
$6.83\times10^{-3}$ & 0.43158 & 0.11865 & $-1.148589$ & $-1.212282$ & $-6.37\times10^{-2}$ & metastable\\
$9.05\times10^{-3}$ & 0.43061 & 0.10696 & $-1.156521$ & $-1.248875$ & $-9.24\times10^{-2}$ & metastable\\
\bottomrule
\end{tabular*}}
\end{table}